\documentclass[journal]{IEEEtran}
\IEEEoverridecommandlockouts
\usepackage{amsmath,amssymb}%,amsfonts}
\usepackage[euler]{textgreek}

\usepackage[ruled,vlined]{algorithm2e} %linesnumbered,lined,boxed,commentsnumbered
\usepackage{mathtools}
\usepackage{mathabx}

\DeclarePairedDelimiter\abs{\lvert}{\rvert}%
\DeclarePairedDelimiter\norm{\lVert}{\rVert}%
\makeatletter
\let\oldabs\abs
\def\abs{\@ifstar{\oldabs}{\oldabs*}}
\let\oldnorm\norm
\def\norm{\@ifstar{\oldnorm}{\oldnorm*}}
\makeatother

\usepackage{graphicx}
\usepackage{subcaption}
\usepackage{textcomp}
\usepackage[dvipsnames]{xcolor}
\usepackage{color, colortbl}
\definecolor{Gray}{gray}{0.9}
\usepackage{tablefootnote}
\usepackage{threeparttable}

\usepackage{tikz}
\usetikzlibrary{shapes, arrows, snakes}
\tikzset{>=latex}
\newlength{\rx}
\newlength{\ry}

\usepackage{pgfplots}
\pgfplotsset{compat=1.16}
\definecolor{colorblue}{rgb}{0.12156862745098,0.466666666666667,0.705882352941177} % blue
\definecolor{colorgreen}{rgb}{0.172549019607843,0.627450980392157,0.172549019607843} % green
\definecolor{colorred}{rgb}{0.83921568627451,0.152941176470588,0.156862745098039} % red
\pgfplotsset{every axis/.append style={
font=\footnotesize,
label style={font=\footnotesize},
tick label style={font=\scriptsize}  
}
}

\usetikzlibrary{pgfplots.groupplots}

\usepackage{filecontents}

\newcommand{\wolf}[0]{{Mr.\,Wolf}}
\newcommand{\eegnet}[0]{EEGNet}
\newcommand{\edgeeegnet}[0]{MI-BMInet}

\newcommand{\physionetmmmi}[0]{MM/MI}
\newcommand{\bcicompivtwoa}[0]{IV-2a}

\usepackage[acronym, style=super, nonumberlist]{glossaries}
\newacronym{eeg}{EEG}{electroencephalogram}
\newacronym{cnn}{CNN}{convolutional neural network}
\newacronym{bmi}{BMI}{brain--machine interface}
\newacronym{bci}{BCI}{Brain--computer interface}
\newacronym{mi}{MI}{motor imagery}
\newacronym{mm}{MM}{motor movement}
\newacronym{smr}{SMR}{sensory motor rhythm}
\newacronym{mcu}{MCU}{microcontroller unit}
\newacronym{cv}{CV}{cross-validation}
\newacronym{lr}{LR}{learning rate}
\newacronym{dr}{DR}{dropout rate}
\newacronym{hp}{HP}{hyperparameter}
\newacronym{sgd}{SGD}{stochastic gradient descent}
\newacronym{dsp}{DSP}{digital signal processing}
\newacronym{fpu}{FPU}{floating-point unit}
\newacronym{lda}{LDA}{linear discriminant analysis}
\newacronym{mops}{MOPS}{million operations per second}
\newacronym{gops}{GOPS}{giga operations per second}
\newacronym{soa}{SoA}{state-of-the-art}
\newacronym{macc}{MACC}{multiply-and-accumulate}

\newacronym{simd}{SIMD}{Single Instruction, Multiple Data}
\newacronym{elu}{ELU}{Exponential Linear Unit}
\newacronym{relu}{ReLU}{Rectified Linear Unit}
\newacronym{rpr}{RPR}{Random Partition Relaxation}
\newacronym{dma}{DMA}{Direct Memory Access}
\newacronym{svm}{SVM}{support vector machine}
\newacronym{svd}{SVD}{Singular Value Decomposition}
\newacronym{evd}{EVD}{Eigendecomposition}
\newacronym{iir}{IIR}{Infinite Impulse Response}
\newacronym{fir}{FIR}{Finite Impulse Response}
\newacronym{fc}{FC}{Fabric Controller}
\newacronym{mrc}{MRC}{Multiscale Riemannian Classifier}
\newacronym{flop}{FLOP}{Floating Point Operation}
\newacronym{sos}{SOS}{Second-Order Section}
\newacronym{ipc}{IPC}{Instructions per Cycle}
\newacronym{tcdm}{TCDM}{Tightly Coupled Data Memory}
\newacronym{fma}{FMA}{Fused Multiply Add}
\newacronym{alu}{ALU}{Arithmetic Logic Unit}
\newacronym{gpu}{GPU}{Graphics Processing Unit}
\newacronym{soc}{SoC}{System-on-Chip}
\newacronym{csp}{CSP}{Common Spatial Patterns}
\newacronym{fbcsp}{FBCSP}{Filter-Bank \acrlong{csp}}
\newacronym{pulp}{PULP}{parallel ultra-low power}
\newacronym{bn}{BN}{Batch Normalization}
\newacronym{isa}{ISA}{Instruction Set Architecture}

\newacronym{iot}{IoT}{Internet-of-Things}
\newacronym{mmmi}{MM/MI}{Motor Movement/Imagery}

\newacronym{ste}{STE}{straight-through estimator}

\newacronym{ml}{ML}{machine learning}
\newacronym{dl}{DL}{deep learning}
\newacronym{adc}{ADC}{analog-to-digital converter}

\newacronym{fpga}{FPGA}{field-programmable gate array}
\newacronym{wola}{WOLA}{weighted overlap-add}

\newacronym{iom}{IoM}{Internet of Minds}
\usepackage{url}
\usepackage{comment}
\usepackage[]{svg} %inkscapearea=page
\usepackage{adjustbox}
\setsvg{svgpath=02_figures/}
\usepackage{pgfplotstable}
\usepackage{multirow}
\usepackage{booktabs} % for tables 

\usepackage{etoolbox}
\makeatletter
\patchcmd{\@makecaption}
  {\scshape}
  {}
  {}
  {}
\makeatother

\makeatletter
\newcommand\notsotiny{\@setfontsize\notsotiny\@vipt\@viipt}
\makeatother

\def\BibTeX{{\rm B\kern-.05em{\sc i\kern-.025em b}\kern-.08em
    T\kern-.1667em\lower.7ex\hbox{E}\kern-.125emX}}
\usepackage{mathtools}

\newcommand{\new}[1]{{\color{black}#1}}

\newcommand{\xia}[1]{{\color{black}#1}}
\newcommand{\lucar}[1]{{\color{black}#1}}

\makeatletter
\def\ps@IEEEtitlepagestyle{%
  \def\@oddfoot{\mycopyrightnotice}%
  \def\@oddhead{\hbox{}\@IEEEheaderstyle\leftmark\hfil\thepage}\relax
  \def\@evenhead{\@IEEEheaderstyle\thepage\hfil\leftmark\hbox{}}\relax
  \def\@evenfoot{}%
}
\def\mycopyrightnotice{%
  \begin{minipage}{\textwidth}
  \centering \scriptsize
  \copyright 2024 IEEE.  Personal use of this material is permitted.  Permission from IEEE must be obtained for all other uses, in any current or future media, including reprinting/republishing this material for advertising or promotional purposes, creating new collective works, for resale or redistribution to servers or lists, or reuse of any copyrighted component of this work in other works. Published version: DOI: 10.1109/JSEN.2024.3353146
  \end{minipage}
}
\makeatother

\begin{document}
\bstctlcite{IEEEexample:BSTcontrol}

\title{MI-BMInet: An Efficient Convolutional Neural Network for Motor Imagery Brain--Machine Interfaces with EEG Channel Selection}

%{XeegNet: An Extremely Efficient Convolutional Neural Network for Motor Imagery Brain--Machine Interfaces with EEG Channel Selection}
% \title{\edgeeegnet{}: Edge Computing for EEG-based Motor Imagery Brain--Machine Interfaces using a Compact Convolutional Neural Network}
%\title{Edge Computing for EEG-based Motor Imagery Brain--Machine Interfaces}
%\title{An End-to-end Framework for Smart Wearable Brain--Machine Interfaces: from Algorithmic Design to Energy-efficient Embedded Edge Computing} % Deployment on the Edge % resource-constrained algorithmic design % embedded computing? % ultra-edge?
% channel selection near-sensor Motor-Imagery Brain--Computer Interface for Low-Power Edge Computing
% An End-to-end Framework for Smart Wearable Sensors

\author{Xiaying~Wang,~\IEEEmembership{Member,~IEEE,}
        Michael~Hersche,~\IEEEmembership{Student Member,~IEEE,}
        Michele~Magno,~\IEEEmembership{Senior Member,~IEEE,}
        Luca~Benini,~\IEEEmembership{Fellow,~IEEE}% <-this % stops a space
\thanks{Manuscript received November 07, 2019; revised January 16, 2020; accepted February 10, 2020.}
\thanks{X. Wang, M. Hersche, and L. Benini are with the Integrated Systems Laboratory, ETH Z{\"u}rich, Switzerland (e-mail: xiaywang@iis.ee.ethz.ch). 
M. Magno is with the Center for Project-Based Learning D-ITET, ETH Z{\"u}rich, Switzerland. 
M. Hersche is also with Cloud and AI Systems Research, IBM Research, Zürich, Switzerland. 
L. Benini is also with the Department of Electrical, Electronic and Information Engineering, University of Bologna, Italy.} 
\thanks{This project was supported by the Swiss Data Science Center PhD Fellowship under grant ID P18-04.}
%\thanks{Copyright (c) 2020 IEEE. Personal use of this material is permitted. However, permission to use this material for any other purposes must be obtained from the IEEE by sending a request to pubs-permissions@ieee.org.}
    }

\maketitle

\begin{abstract}
A brain--machine interface (BMI) based on motor imagery (MI) enables the control of devices using brain signals while the subject imagines performing a movement. It plays a vital role in prosthesis control and motor rehabilitation. 
\new{To improve user comfort, preserve data privacy, and reduce the system's latency, a new trend in wearable BMIs is to execute algorithms on low-power microcontroller units (MCUs) embedded on edge devices to process the electroencephalographic (EEG) data in real-time close to the sensors.}
However, most of the classification models present in the literature are too resource-demanding, making them unfit for low-power MCUs.
This paper proposes an efficient convolutional neural network (CNN) for EEG-based MI classification that achieves comparable accuracy while being orders of magnitude less resource-demanding and significantly more energy-efficient than state-of-the-art (SoA) models for a long-lifetime battery operation. To further reduce the model complexity, we propose an automatic channel selection method based on spatial filters and quantize both weights and activations to 8-bit precision with negligible accuracy loss. \new{Finally, we implement and evaluate the proposed models on leading-edge parallel ultra-low-power (PULP) MCUs. The final 2-class solution consumes \lucar{as little as 30\,\textmu J/inference with a runtime of 2.95\,ms/inference}} and an accuracy of 82.51\% while using 6.4$\times$ fewer EEG channels, becoming the new SoA for embedded MI-BMI and defining a new Pareto frontier in the three-way trade-off among accuracy, resource cost, and power usage.
\end{abstract}

% with the drawback of diminishing the number of features and the numerical precision that can lead to accuracy degradation. 
% Techniques such as feature selection and quantization help reducing the resource requirements, but might lead to accuracy degradation. 
% to tackle the inter-session and inter-subject challenges


\section{Introduction}

% what is bmi
A \gls{bmi} aims to translate brain activities into actionable information to control external devices, such as a wheelchair~\cite{xiong2019low} or a prosthesis~\cite{VILELA2020prosthesis}.
Besides clinical relevance for patients to regain lost abilities~\cite{naturereview2016tubingenRehab}, recent developments in wearable technologies have pushed the field of \gls{bmi} towards everyday life tasks in consumer products~\cite{casson2019wearableeeg,arico2020bcidailylife}, for example, to control robots~\cite{shao2020cleaningrobot} or drones~\cite{kobayashi2019ssvepdrone}, yielding improved user experience also for healthy subjects~\cite{belkacem2020bci4lifequality}.
\Gls{mi} is of great interest among the current \gls{bmi} paradigms because it does not strictly depend on external stimuli; hence, it can be independently and asynchronously self-paced~\cite{Freer2020_selfpacedMI}. The \gls{mi}-\gls{bmi} decodes the user's intention by analyzing the brain activities while the subject thinks of a movement of a body part without actually performing it~\cite{Pfurtscheller1999Event-relatedPrinciples}.

\new{A recent work by Zhuang et al.~\cite{Zhuang2021tsmc} has demonstrated the successful usage of \gls{mi}-\gls{bmi} in controlling ground vehicles using a wireless wearable device based on non-invasive \glspl{eeg}. 
%Despite the high merit of the paper, t
The data processing and the classification are executed on a remote engine, which leads to potential privacy concerns, a longer latency, and higher power consumption due to the data transmission.
Recent achievements in low-power processing platforms and miniaturization enable the sensor data processing directly at the edge~\cite{Kartsch2019BioWolf:Connectivity_short,Schneider2020}.
By embedding algorithms at the sensor node, the data is analyzed locally, preserving privacy, reducing the energy consumption for a longer battery lifetime, and minimizing the system's latency~\cite{beach2021edgealgorithmswearables}. %, otherwise caused by remote processing.
The decoded information can be directly sent to the controlled devices without any intermediate apparatus between the brain and the machines, empowering \lucar{the future \gls{iom}~\cite{Tunstel2021_futureDirections}.}}

To achieve this goal, a smart wearable \gls{bmi} has to satisfy a three-way trade-off among
a) algorithmic performance: the embedded algorithm has to be able to complete the targeted task accurately; b) computational and storage parsimony: the complexity of the deployed algorithm has to meet the resource constraints of the edge platform in terms of memory footprint and runtime for a real-time application scenario; c) power: the power consumption must be low to guarantee a long-term continuous operation with long battery lifetime~\cite{beach2021edgealgorithmswearables}.
The low signal-to-noise ratio of the \gls{eeg} data and the inter-session and inter-subject variability pose enormous challenges to obtaining high classification accuracy. Several \gls{ml} and \gls{dl} models have been proposed in the literature, achieving remarkable algorithmic performance~\cite{Schirrmeister2017DeepVisualization,wu2019_MSFBCNN,bang2021fb3dcnn}. However, the majority of them target only accuracy as a key metric while ignoring the resources required by the model, making them unfit for low-power \glspl{mcu}. 
%
%For example, the leading edge \gls{pulp} platform based on the open-source RISC-V \gls{isa} has proven to be at least one order of magnitude more energy-efficient than the most popular commercial ARM Cortex-M series~\cite{Wang_fann-on-mcu,Schneider2020}. 
For example, the Mr. Wolf processor~\cite{pullini2019wolf} from the \gls{pulp} platform, a leading-edge example of high-performance and low-power \gls{mcu}, has less than 600\,kB of on-chip fast memory and can deliver up to 16.4 \Gls{gops} of computational capability. \new{It has been embedded in a \lucar{wearable \gls{bmi} device}, called Biowolf, \lucar{with the minimal form factor of 40\,mm$\times$20\,mm$\times$2\,mm and} consuming less than 10\,mW~\cite{Kartsch2019BioWolf:Connectivity_short}.
When considering very low-power \glspl{mcu} in the sub-100-milliwatt power range, almost all the \gls{soa} networks cannot be deployed on-board without additional expensive off-chip memory, yielding increased power consumption and longer execution latency. In full-precision representation, even the most compact EEGNet~\cite{Lawhern2018EEGNet:Interfaces} is out-of-reach by requiring more than 1\,MB for storing the two biggest consecutive feature maps with the standard layer-by-layer execution during inference, making the deployment more challenging~\cite{Wang2020_memea,Schneider2020}.
In contrast, edge devices with more resources are too power-hungry and \lucar{do not} meet the specifications for long-term usage.}
%It is specifically designed for edge processing, but even the EEGNet~\cite{Lawhern2018EEGNet:Interfaces}, which is well-known for its compact model size, is out of reach for embedding on this device by requiring more than 1\,MB of memory to store intermediate results during inference computation, \new{making the deployment more challenging~\cite{Wang2020_memea,Schneider2020}.}
Therefore, resource-friendly models that at the same time maintain high accuracy are desirable.
%Therefore, smaller models which demand fewer resources but at the same time maintain comparable accuracy are desirable. %~\cite{Li2020MicroNetTI}. 
%The resource requirements need to be considered already at the model design stage.

\new{Besides considering the resource usage already during the model design, additional methods to reduce the model complexity are essential.
In \gls{eeg} applications, an important topic is the channel selection, where the \gls{eeg} channels with more relevant features are selected for the final classification stage~\cite{das2015smcconf, Tokovarov2020}.
%Often, it is motivated for an improved user comfort, but in this work we view it also as an effective technique to reduce the memory 
In this work, we emphasize the importance of this technique not only for selecting the most significant features to improve accuracy but also for its advantages of reducing memory usage and computational complexity. Moreover, \lucar{the reduction of \gls{eeg} channels lowers the system's power consumption since fewer analog front-end circuits are necessary for data acquisition, and it is fundamental for improving user comfort and achieving optimal device wearability.}
%Feature selection is an effective technique to reduce the number of features, yielding a smaller memory footprint and lower computational complexity. Especially for \gls{eeg} applications, decreasing the number of \gls{eeg} channels brings the benefit of not only a reduced number of computations but also lower power consumption since fewer analog front-end circuits are necessary for the data acquisition. However, a reduced number of channels makes the task even more difficult, often causing accuracy degradation~\cite{Wang2020_memea}. Hence, sophisticated methods for channel selection are necessary to extract relevant features for the task while maintaining accurate performance~\cite{das2015smcconf, chen2020csp, Tokovarov2020}.
%
Another effective method to reduce the resource requirements is quantization, e.g., using 8 bits to represent numbers instead of 32 bits~\cite{quantlab2019,Cavigelli2020RPR:Networks}. It yields a significant reduction in memory usage and allows the use of efficient hardware units. 
Worth to be noted that both channel reduction and quantization {pose challenges} to the classification task and can lead to accuracy degradation due to the reduced number of inputs and numerical precision~\cite{Wang2020_memea}. Hence, sophisticated algorithms are demanded to guarantee comparable accuracy while using minimal resources.
%However, reduced numerical precision can lead to accuracy loss; hence, sophisticated algorithms are necessary to guarantee comparable performance~\cite{Cavigelli2020RPR:Networks,Schneider2020}.
}

\begin{figure}[t]
    \centering
    \includegraphics[width=\columnwidth]{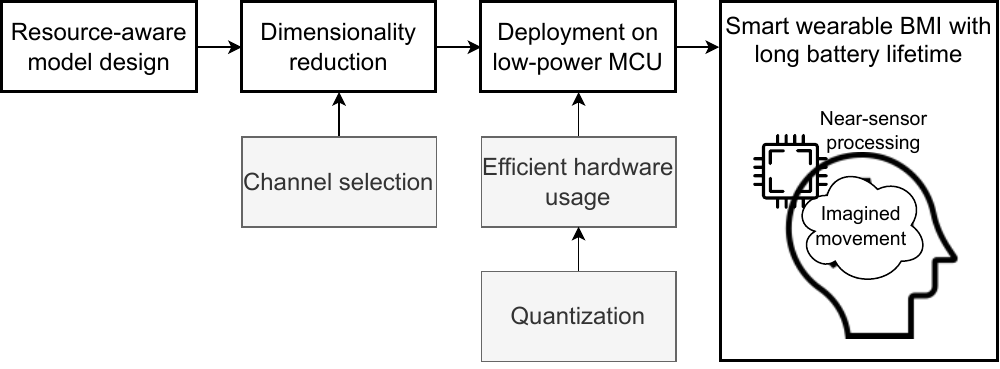}
    \caption{\new{Overview of the proposed methods for a smart wearable \gls{mi}-\gls{bmi} with embedded near-sensor processing.}}
    \label{fig:overview}
\end{figure}

\new{In this paper, we propose an end-to-end workflow for the realization of an energy-efficient wearable \gls{mi}-\gls{bmi} enabled with smart near-sensor computing, as shown in Fig.~\ref{fig:overview}.
We select \glspl{mcu} based on \gls{pulp} platforms~\cite{pullini2019wolf,Rossi2021vega}, because they have been demonstrated to outperform other families of low-power \glspl{mcu} thanks to the parallel cores and the custom hardware extensions for \gls{dsp}~\cite{Wang_fann-on-mcu,Garofalo2019PULP-NN:Processors,Schneider2020,Wang2021mrc_tbiocas}.
%First, we co-design a compact, yet accurate classification model based on the available hardware resources. Second, we reduce the input dimensions by automatically selecting the most relevant \gls{eeg} channels. Third, we further reduce the memory usage by quantizing both model weights and features to 8-bit precision without significant accuracy degradation. Finally, we deploy the quantized model on the selected \glspl{mcu} by efficiently exploiting the underlying hardware architecture.
Our main contributions are as follows:
\begin{itemize}
    \item We co-design a compact yet accurate \gls{cnn} for \gls{mi}-\gls{bmi} classification constrained to minimal hardware resources. It meets the tight resource limitations of ultra-low-power \glspl{mcu} while being as accurate as the resource-demanding \gls{soa} algorithms. We evaluate the performance in both inter-session and inter-subject challenges using two publicly available \gls{mi} datasets, namely the BCI Competition IV-2a and the Physionet EEG \gls{mmmi}, and achieve respectively 76.03\%, 65.62\% accuracy on the {4-class} task and 86.32\%, 82.79\% on the {2-class} task.
    \item We reduce the input dimensions by automatically selecting the most relevant \gls{eeg} channels to limit further the model complexity and the memory footprint while maintaining similar classification accuracy (82.79\% vs. 82.51\% for the {2-class} task of the \physionetmmmi{} dataset) or even improving the accuracy by up to 1.33\% in the case of the subject-specific models of the \bcicompivtwoa{} dataset. %based on the spatial filters of the proposed \gls{cnn}
    \item We quantize both model weights and activations to 8-bit precision with negligible effect on the classification accuracy to further reduce the memory usage and optimally exploit the hardware extensions. %(less than 0.4\% drop)
    %We further decrease the model size by quantizing both weights and activations of the proposed \gls{cnn} to 8-bit representations with negligible effect on the classification accuracy.
    We efficiently deploy the quantized models on ultra-low-power \glspl{mcu}, achieving an energy consumption of merely 30\,\textmu J per inference and real-time execution of 2.95\,ms. With a 6.4$\times$ reduction of \gls{eeg} channels, the battery operation time can be increased from 3.8 up to 22 hours.
    %and experimentally measure the power consumption and the model execution runtime. 
    %Our best solution for the {2-class} task of the \physionetmmmi{} dataset takes only 2.95\,ms and consumes 30\,\textmu J per inference, using only 10 \gls{eeg} channels instead of 64.
\end{itemize}
This work successfully satisfies the three-way trade-off among performance, cost, and power~\cite{beach2021edgealgorithmswearables} by maximizing the algorithmic accuracy, minimizing the computational cost, and minimizing the power consumption for a real-time smart wearable \gls{mi}-\gls{bmi} with an extended battery lifetime, paving the way to the future \gls{iom}\footnote{Open-source code: \url{https://github.com/pulp-platform/MI-BMInet}.}. % for long-term usage. %This is an crucial step towards the future \gls{iom}. paving the way to the future \gls{iom}.
%We release open-source code\footnote{It will be released after review.}.
}

\section{Related Works}\label{sec:related}

%First of all, we discuss the \gls{soa} algorithms for \gls{mi}-\gls{bmi} classification, followed by feature/channel selection methods. We then report quantization and embedded deployment techniques together with an overview on low-power \glspl{mcu}.
\new{
We discuss the related works based on the popular BCI Competition \bcicompivtwoa{} dataset~\cite{Brunner2008BCIA,tangermann2012_bcicompreview} and the Physionet EEG \gls{mmmi} dataset~\cite{Schalk2004BCI2000:System,goldberger2000physiobank}. The former is the most representative for subject-specific models and inter-session challenges, whereas the latter includes a remarkable number of subjects and is often used to tackle the inter-subject variability. We select only the works that follow the competition rules for the \bcicompivtwoa{} dataset or
%adopt the validation methodology that tackles the inter-subject challenges by training 
train a global model validated on unseen subjects for the \physionetmmmi{} dataset for a correct and fair comparison.
As this paper covers the full flow of model design, channel selection, and embedded implementation, we divide the discussion of the related works accordingly.
}

\new{
\subsection{Classification Models}\label{subsec:class_models}

Traditional \gls{ml}-based approaches require manual feature extraction, based on which a classifier is trained to solve the task.
%The \gls{mi}-\gls{bmi} literature can be divided into two classification approaches:
%Two classification approaches are currently present in \gls{mi}-\gls{bmi} literature: 
%a) traditional \gls{ml} with manual feature engineering; b) end-to-end \gls{dl} with automatic feature extraction. 
The most common and effective methods to extract discriminative features in \gls{mi}-\glspl{bmi} are based on the \gls{csp}~\cite{ang2012fbcspbcicomdatasets,chen2020csp} and the Riemannian geometry~\cite{Yger2017RiemannianReview}. 
The authors in~\cite{das2015smcconf} and \cite{Jiang2020_csp} propose spatial or temporal filtering based on \gls{csp}, reaching a {2-class} classification accuracy of 80.56\% and 79.6\% on the \bcicompivtwoa{} dataset, respectively.
%Jiang et al.~\cite{Jiang2020_csp} propose spatio-temporal filtering based on \gls{csp} reaching a {2-class} classification accuracy of 79.6\% on the \bcicompivtwoa{} dataset.
%
The work in~\cite{Hersche2018FastFeatures} compares the \gls{csp} and Riemannian methods and demonstrates that \gls{mrc} achieves the best {4-class} classification accuracy of 75.47\% using multiple time windows with a \gls{svm}. 
A more recent work~\cite{Wang2021mrc_tbiocas} based on \gls{mrc} reduces the number of features and achieves a {4-class} accuracy of 76.4\% using different classifiers for each subject.
By combining \gls{fbcsp}, autoregressive models, and feature selection with mutual information, the work in~\cite{Wang2020_mi-ts} achieves the highest 2-class accuracy of 86.01\%. The authors generalize the methods to the 4-class task obtaining a Cohen's kappa coefficient of 0.61. Similar techniques of feature extraction are used in~\cite{das2020fbcspovr} with Adaptive Boosting classifier, resulting in an improved kappa value of 0.646. %However, the same methods perform worse in the 4-class task than~\cite{das2020fbcspovr} reaching a kappa value of 0.61 instead of 0.65.

In another vein, \gls{dl} methods enable the classification of \gls{bmi} tasks without handcrafted features. Many works have demonstrated the effectiveness of \glspl{cnn} in learning temporal and spatial features obtaining accuracy values ranging from 79.9\% up to 86.96\% for {2-class} tasks~\cite{Schirrmeister2017DeepVisualization,bang2021fb3dcnn} and 74.31\% to 77.35\% for {4-class} tasks~\cite{wu2019_MSFBCNN,Amin2019_MCNN,ingolfsson2020eegtcnet,Schneider2020,Salami2022_eegitnet}. With additional hyperparameters tuning specific to each subject, the accuracy can be improved up to 83.84\% on the {4-class} task~\cite{ingolfsson2020eegtcnet}. For the inter-subject cases on \physionetmmmi{} dataset, the {2-} and {4-class} accuracy values range from 80.38\% to 83.26\% and 58.58\% and 65.07\%, respectively~\cite{Dose2018AnBCIs,Wang2020_memea,Tokovarov2020}.
One major drawback of the \gls{dl} approach is that the models tend to grow in size, increasing the demand for computational and storage resources. 
FB3DCNN achieves the \gls{soa} accuracy (86.96\%) in the 2-class task of the \bcicompivtwoa{} dataset, but it requires storage for 46 million parameters~\cite{bang2021fb3dcnn,ingolfsson2020eegtcnet}.
EEG-TCNet and EEG-ITNet perform best in the 4-class task (77.35\% and 76.74\%, respectively) and have significantly fewer parameters (below 5\,k)~\cite{ingolfsson2020eegtcnet,Salami2022_eegitnet}. However, when considering the standard layer-by-layer computation schedule, the additional memory footprint required during the execution time is up to two orders of magnitude more than the number of parameters, making them significantly more challenging for very low-power \glspl{mcu}.
%
%When considering very low-power \glspl{mcu} in the sub-100-milliwat power range, e.g., Mr. Wolf, almost all the \gls{soa} networks, reported in Table~\ref{tab:background}, are impossible to be implemented on-board without additional expensive off-chip memory, yielding increased power consumption and longer execution latency. In full-precision, even the most compact EEGNet is out-of-reach by requiring more than 1\,MB for storing two biggest consecutive feature maps with the common layer-by-layer execution.
%In contrast, edge devices with more resources are too power-hungry and might not meet the specifications for long-term usage.
%
In this paper, we propose a more compact \gls{cnn} with less than 50\,k parameters \textit{and features} to be stored during the inference time. The resource requirement is orders of magnitude lower than recent related works~\cite{wu2019_MSFBCNN,Amin2019_MCNN,ingolfsson2020eegtcnet,bang2021fb3dcnn} while achieving similar classification accuracy.
%the accuracy values of 86.32\% and 76.03\% for 2- and 4-class tasks of \bcicompivtwoa{} dataset, respectively. 
%(Note that our accuracy values are averaged over multiple runs to account for variability due to the nondeterministic behaviour of \gls{cnn} training for a better statistical significance. Whereas, the accuracy reported in~\cite{ingolfsson2020eegtcnet} is the maximum value over multiple runs to obtain a better result.)
%with up to 9.8$\times$ reduction in the memory usage for the feature maps compared to EEG-TCNet, while achieving an accuracy of  being up to 4.73\% more accurate on the \bcicompivtwoa{} dataset.
}

\subsection{Channel Selection}

%Additionally, feature selection allows to reduce the number of features and to identify the most relevant ones, usually yielding improved performance and fewer computations. In \gls{eeg}-based applications, it is beneficial to reduce the number of \gls{eeg} channels directly. Using fewer electrodes means decreased power consumption for data acquisition and computation, improved user comfort, and reduced setup time for the acquisition headset. 
\new{It is beneficial to reduce the number of \gls{eeg} channels for improved user comfort, a reduced setup time, a lower power consumption, and a decreased model complexity.}
%Subject-specific channel selection has proven to be advantageous also for increasing classification accuracy~\cite{das2015smcconf,chen2020cssimilarity}.
\new{
%\gls{csp}-based methods have proven to be effective. 
The authors in~\cite{das2015smcconf} have proven that input \gls{eeg} channels can be effectively reduced subject-specifically from 22 to an average of 8.11 while improving the 2-class accuracy on the \bcicompivtwoa{} dataset by around 3\% using \gls{csp}-based methods.
%On the 2-class task of the \bcicompivtwoa{} dataset, the \gls{eeg} channels can be reduced subject-specifically from 22 to an average of 8.11 while improving the accuracy by around 3\%~\cite{das2015smcconf}. 
A more recent work~\cite{Gaur2019_memdbf} uses a fixed number of 15 channels for all subjects reaching an accuracy of 79.19\%. It is 4.42\% less than~\cite{das2015smcconf}, demonstrating that selecting a variable number of channels for each subject is beneficial.
}
The authors in~\cite{chen2020cssimilarity} combine \gls{csp} with Riemannian distances obtaining an \gls{soa} classification accuracy of 77.82\% and a kappa value of 0.71 with an average of 15.2 selected channels over the nine subjects of the \bcicompivtwoa{} dataset for the 4-class \gls{mi} task. \gls{cnn}-based channel reduction approaches are also found in the literature~\cite{Dose2018AnBCIs,Tokovarov2020}. Dose et al.~\cite{Dose2018AnBCIs} manually select subsets of \gls{eeg} channels to compare with related works and demonstrate that their proposed architecture based on the shallow ConvNet~\cite{Schirrmeister2017DeepVisualization} outperforms most other models. More recent work by Tokovarov et al.~\cite{Tokovarov2020} applies an automatic channel selection method based on \gls{cnn} feature maps obtaining 82.34\% accuracy with only 14 instead of 64 channels on the 2-class \gls{mi} task of the \physionetmmmi{} dataset outperforming the manual selection of~\cite{Dose2018AnBCIs}.
Given the many advantages of channel selection, in this work, we propose an automatic method based on the spatial filters of the proposed \gls{cnn}. It can effectively reduce the number of channels from 64 to 10 for the \physionetmmmi{} dataset, with the advantage of 1.3$\times$ fewer parameters, 3.1$\times$ less memory footprint, and 1.4$\times$ lower computational complexity while having a negligible accuracy drop of 0.28\% for the 2-class inter-subject task. The inter-session accuracy of the subject-specific models for the \bcicompivtwoa{} dataset is increased by up to 1.33\%, with an average channel reduction of 60\% for the 2-class task.

\subsection{Embedded Implementation}

Over the recent years, increasing attention has been gained by the embedded deployment of \gls{ml} and \gls{dl} models on low-power edge devices, nourishing the fast-growing field of TinyML~\cite{tinyML} and giving birth to notable projects such as TensorFlow Lite~\cite{david2021tensorflow}.
%Embedded deployment of \gls{ml} and \gls{dl} models on low-power edge devices has been gaining increasing attention over the recent years, nourishing the fast-growing field of TinyML~\cite{tinyML} and giving birth to notable projects such as TensorFlow Lite~\cite{david2021tensorflow}.
%giving birth to notable projects and forming new research community such as TensorFlow Lite~\cite{david2021tensorflow} and TinyML~\cite{tinyML}. %giving birth to famous projects such as TensorFlow Lite and forming new research community such as TinyML. 
Some initial efforts can be also found in the \gls{bmi} literature.
%However, most of the \gls{bmi} algorithms, especially the ones for \gls{mi}, are too resource-demanding for low-power \glspl{mcu}. In contrast, edge devices with more resources are too power-hungry and might not meet the specifications for long-term usage.
%
\new{Belwafi et al.~\cite{belwafi2018_wolacsp} propose a \gls{csp}-based approach implemented in 16-bit precision on a Stratix-IV \gls{fpga} board with a power consumption of 700\,mW. It achieves a 2-class accuracy of 78.85\% on the \bcicompivtwoa{} dataset with an inference time of 430\,ms and energy consumption of \lucar{301,000\,\textmu J.}}
Another embedded implementation based on \gls{csp} can be found in~\cite{malekmohammadi2019_cspsvm}. 
%The authors employ a \gls{csp}-based feature extractor, mutual information as feature selector, \gls{lda} to reduce features, and \gls{svm} for the final classification. 
The accuracy on the \bcicompivtwoa{} dataset is 80.55\% and 67.21\% for \mbox{2-} and 4-class tasks, respectively. The methods are implemented in 20-bit precision on a Virtex-6 \gls{fpga} consuming 83.90\,mW and taking up to 11.66\,ms \new{(i.e., 978\,\textmu J).} % in the worst case.
%The authors acquire an own dataset of left and right \gls{mi} tasks using 14 channels and implement the proposed methods on a Virtex-6 \gls{fpga} consuming 83.90\,mW.

Two other works have implemented \gls{dl}-based approaches for \gls{mi}-\gls{bmi} on embedded devices~\cite{Wang2020_memea,Schneider2020}. 
\new{The authors in~\cite{Wang2020_memea} deploy an \eegnet{}-based \gls{cnn} on two STMicroelectronics \glspl{mcu} featuring ARM Cortex-M4 and M7. %, yielding a power consumption of 42.44\,mW and up to 413.06\,mW, respectively. 
%The proposed \gls{cnn} in~\cite{Wang2020_memea} based on \eegnet{} outperforms Dose et al.~\cite{Dose2018AnBCIs}. The selected \glspl{mcu} are based on ARM Cortex-M4 and M7 consuming respectively 42.44\,mW and up to 413.06\,mW. 
The model requires too much memory for the selected low-power \glspl{mcu} based on the X-CUBE-AI deployment tool provided by the manufacturer.
Hence, the input signals are downsampled, shortened in time, and 38 manually selected \gls{eeg} channels are used. % to deploy the models within the constrained resources. 
The 4-class accuracy on the \physionetmmmi{} dataset drops by 2.56\% on the Cortex-M4 due to the limited hardware resources. The embedded model with the highest accuracy (64.76\%) could be implemented only on the Cortex-M7, which provides more resources but consumes one order of magnitude more power than the Cortex-M4. The model with the highest accuracy takes 43.81\,ms and consumes \lucar{18,100\,\textmu J} per inference. 
%The accuracy for the 4-class \gls{mi} task drops negligibly from 65.07\% to 64.76\% on the Cortex-M7 which has more resources but consumes one order of magnitude more power than the Cortex-M4. Whereas, the decrease in accuracy on the Cortex-M4 is more accentuated (-2.56\%). The inference time is around 100\,ms on the M4 and down to 20.40\,ms on the M7 for the same model.
% oth Model 1 (62.51% accuracy)and Model 2 (64.76%) use a downsampling ofds=3 andNch=38 channels,the former has time windowT=1 s, while the latterT=2 s. 64 channels 65.07%
A more energy-efficient implementation is proposed in~\cite{Schneider2020}, outperforming all previous solutions based on \glspl{fpga} and Cortex-M \glspl{mcu}.
The model is first quantized to 8-bit fixed-point representation with negligible loss in accuracy and implemented on \wolf{}, consuming only 11.75\,mW.} %, a RISC-V-based \gls{pulp} platform with parallel processing units~\cite{pullini2019wolf}. 
The average quantized accuracy over the nine subjects of the \bcicompivtwoa{} dataset is 70.9\% for the 4-class \gls{mi} task. The energy consumption is 337\,\textmu J, and the inference time is 28.67\,ms. 
\new{However, the classification accuracy is relatively low, and the proposed solution is not flexible because it is designed specifically for \eegnet{} and it abandons the common layer-by-layer paradigm adopted by most deployment frameworks~\cite{tinyML}.}
%Moreover, no channel reduction has been explored to further reduce the complexity and improve the energy-efficiency. 

\new{A very recent work~\cite{Wang2021mrc_tbiocas} proposes an \gls{mrc} implementation with a combination of 8-, 16-, and 32-bit fixed-point and 32-bit floating-point representations on the new \gls{pulp} \gls{mcu} called Vega~\cite{Rossi2021vega}. It is currently the \gls{soa} in terms of accuracy and energy efficiency (74.1\% 4-class quantized accuracy, 16.9\,ms inference time, and 198\,\textmu J energy consumption).}

\new{
In this work, we quantize the proposed \gls{cnn} to 8 bits achieving a 4-class quantized accuracy of 75.63\%. The full-channel implementations for the \bcicompivtwoa{} dataset take 11.37\,ms and 5.10\,ms and consume 114\,\textmu J and 60.5\,\textmu J on Mr. Wolf and Vega, respectively. It is 4.7\% and 1.5\% more accurate than \cite{Schneider2020} and the \gls{soa}~\cite{Wang2021mrc_tbiocas}. The inference execution is 2.5$\times$ and 3.3$\times$ faster, and it consumes 3$\times$ and 3.3$\times$ less on the same \glspl{mcu}, respectively.
The energy consumption is further lowered by reducing the number of \gls{eeg} channels without significantly affecting the accuracy. 
%Our best solutions after the channel selection achieve a lower energy consumption of 54\,\textmu J and 30\,\textmu J, respectively for \bcicompivtwoa{} and \physionetmmmi{} datasets.
}

\section{Methods}\label{sec:methods}

%second, an accurate classification model is designed by taking into consideration the hardware constraints; third, feature selection is performed to reduce the number of \gls{eeg} channels yielding more energy-efficient models; fourth, the model is quantized to further minimize the resource usage; finally, the quantized model is efficiently implemented on a low-power \gls{mcu}. The following sections are organized accordingly.

\new{The workflow proposed in this paper is as follows: first, an accurate classification model is designed by taking into consideration hardware constraints; second, \gls{eeg} channel selection is performed to reduce the input dimensionality yielding more energy-efficient models; third, the model is quantized to minimize further the resource usage and is efficiently deployed on the selected \glspl{mcu} by exploiting the hardware extensions.}

\subsection{Network Architecture and Resource Requirements}\label{sec:net_archi}
% In this section, we describe the novel network architecture proposed in this work. 
% %
% The network is designed to be light-weighted, yet accurate, in order to perform \gls{eeg}-based \gls{mi}-\glspl{bci} using low-power \glspl{mcu} on smart edge wearable devices. Hence we name it \edgeeegnet{}. 

%\new{This section describes the proposed \edgeeegnet{}, a light-weighted yet accurate \gls{cnn} that is suitable for low-power \gls{mi}-\glspl{bmi}.}
% based on \gls{eeg}. %  for \gls{eeg}-based \gls{mi}-\glspl{bmi} that is suitable for low-power \glspl{mcu}. % on smart wearable devices.
\new{As motivated in the introduction, our target is low-power \gls{mi}-\glspl{bmi}, meaning that the classification model has to respect strict resource availability. Hence, while designing the network, we also assess the memory footprint and the computational complexity. Additionally, unlike most previous works, which consider only the number of parameters, we propose a more accurate estimation of the memory usage based on the standard layer-by-layer computation schedule. This is especially crucial for embedded deployment~\cite{lai2018_arm_embedded}.
}

%The network is designed to have variable input channels and is suitable for \gls{eeg} channel selection and reduction. 
%
%In contrast to related \glspl{cnn}~\cite{Dose2018AnBCIs,Lawhern2018EEGNet:Interfaces}, \edgeeegnet{} reduces the spatio-temporal \gls{eeg} signals first in \emph{spatial domain} instead of temporal domain, effectively relaxing the computational and memory requirements of the network.

% \begin{figure*}[ht!]
%     \fontsize{8}{10}\selectfont
%     \centering
%     \includesvg[width=\textwidth]{eegnet}
%     \caption{EEGNet~\cite{Lawhern2018EEGNet:Interfaces} in standard configuration for 4-class MI on the Physionet Motor Movement/Imagery Dataset. A window of 3\,s ($N_s = 480$ samples) with $N_{ch} = 64$ channels is classified at the time.  
%     }
%     \label{fig:cnn}
% \end{figure*}

\begin{figure*}[ht!]
\centering
    %\fontsize{8}{10}\selectfont
    %\includesvg[width=1.12\textwidth]{02_figures/edgeEEGNet_arch_sizes_quant}
    \includegraphics[width=\linewidth]{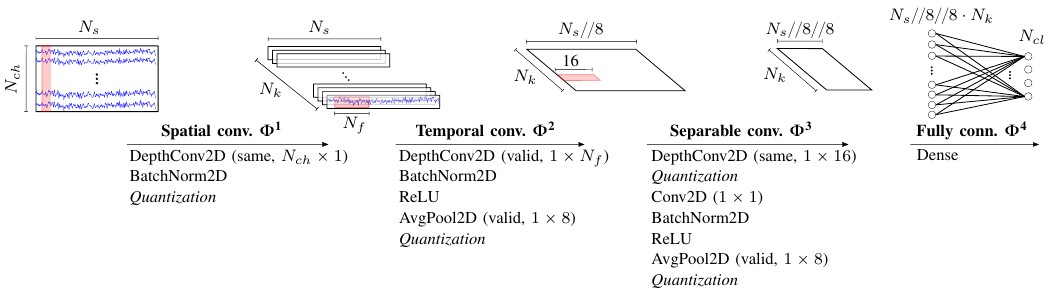}
    \caption{Architecture of \edgeeegnet{}.
    $N_s$ is the number of input samples in the time domain, $N_{ch}$ the number of EEG channels, $N_{cl}$ the number of classes, $N_f$ the filter size of the temporal filter, and $N_k$ is the number of filters. %, and $N_p$ the pooling length.
    %$n_S$ and $n_T$ are respectively the number of filters in spatial and temporal convolution. 
    The padding strategy and the kernel size are reported in brackets. 
    %The standard configuration for 4-class MI on the Physionet EEG \gls{mmmi} Dataset is shown. A window of 3\,s ($N_s = 480$ samples) with $N_{ch} = 64$ channels is classified at a time. The network weights used for EEG channel selection are displayed in red and framed in green.
    The addition of the quantization layers is highlighted in \textit{italic}.}
    \label{fig:cnn}
\end{figure*}

\new{Fig.~\ref{fig:cnn} depicts the proposed model named \edgeeegnet{}.} The input array is arranged such that each row represents the \gls{eeg} signals from the electrodes, and each column is the sample at every time point. The dimension is $N_{ch} \times N_s$, with $N_{ch}$ being the number of \gls{eeg} channels and $N_s$ the number of samples.
\new{
%\gls{eeg} contains the spatio-temporal information of the brain activity. Hence, 
Many works in literature have demonstrated that temporal and spatial features most successfully \lucar{decode} the \gls{eeg} signals, as they represent the temporal and spatial summation of brain activities~\cite{Schirrmeister2017DeepVisualization,Lawhern2018EEGNet:Interfaces,Amin2019_MCNN,ingolfsson2020eegtcnet,Salami2022_eegitnet}. In all previous works, the extraction of the temporal information causes a significant increase in memory usage and computations, leading to deployment difficulties~\cite{Wang2020_memea,Schneider2020}. In our proposed \gls{cnn}, we also extract temporal and spatial features, as \lucar{this is currently the most effective approach} for \gls{eeg} signals, but, in contrast to previous works, we reduce the dimensionality by first extracting the spatial features, followed by the extraction of the temporal information. This way, the memory and the computation overhead are effectively reduced.}

More precisely, the input is passed to a first depthwise convolutional layer with $N_k$ one-dimensional kernels of size $N_{ch} \times 1$ to find spatial correlations among the \gls{eeg} channels, followed by a batch normalization layer. 
%The depth multiplier $d$ is $n_S$ and corresponds to the number of filters and the number of output feature maps of dimension $1\times N_s$.
This spatial convolution outputs $N_k$ feature maps of dimension $1\times N_s$. 
These are subsequently filtered using another layer of depthwise convolution with 
%$d=1$ and 
$N_k$ kernels of dimension $1 \times N_f$ applied along the temporal dimension to learn temporal information. 
\new{The number of parameters in a depthwise convolutional layer can be calculated as $N_k \cdot (k_H \cdot k_W)$,
with $k_H$ and $k_W$ being the height and width of the kernels, reported in brackets in Fig.~\ref{fig:cnn}.}
% \begin{equation}\label{eq:params_cnn}
%     \hash\text{params} =  n \cdot k,
% \end{equation}
% where $n$ is the number of filters and $k$ the kernel size. 
\new{Additionally, each batch normalization layer has $4\cdot f$ parameters, with $f$ being the number of feature maps, which corresponds to the number of filters $N_k$ in the case of depthwise convolutions.}
Afterward, a separable convolution, i.e., a depthwise convolution followed by a pointwise convolution, is applied to extract additional spatio-temporal features using $N_k$ filters of size $1\times 16$. 
\new{This amounts to $N_k \cdot (1 \cdot 16) + N_k \cdot N_k$ parameters.} 
\Gls{relu} is chosen as activation since it is the most hardware-friendly non-linear activation function. Additionally, average pooling layers with a kernel size of $8\times 1$ are applied in the time domain to reduce the dimensions of the feature maps. Finally, a fully connected layer computes the classification output. %A detailed breakdown of the layers is reported in Table~\ref{tab:cnn}.
\new{It presents $(N_{in} +1) \cdot N_{out}$ parameters, with $N_{in}$ being the number of input nodes and $N_{out}$ the output nodes, i.e., $N_{cl}$.}

%\subsection{Resource Requirements}

% \edgeeegnet{} contains two depthwise and one separable convolutions, each with $N_k$ filters.
% %, that is a depthwise convolution followed by a pointwise convolution. 
% The number of parameters in a depthwise convolutional layer can be calculated as $N_k \cdot (k_H \cdot k_W)$,
% with $k_H$ and $k_W$ being respectively the height and width of the kernels, reported in brackets in Fig.~\ref{fig:cnn}.
% % \begin{equation}\label{eq:params_cnn}
% %     \hash\text{params} =  n \cdot k,
% % \end{equation}
% % where $n$ is the number of filters and $k$ the kernel size. 
% While in the separable convolutional layer there are $N_k \cdot (1 \cdot 16) + N_k \cdot N_k$ parameters.
% % \begin{equation}\label{eq:params_cnn}
% %     \hash\text{params} = n \cdot k + n \cdot n.
% % \end{equation}
% Additionally, each batch normalization layer has $4\cdot f$ parameters, with $f$ being the number of feature maps, which corresponds to the number of filters $N_k$ in the case of depthwise convolutions.
% Finally, a fully connected layer with $n_{in}$ input nodes and $n_{out}$ output nodes presents $(n_{in} +1) \cdot n_{out}$ parameters. 
%
Besides the network weights, the input and the output feature maps need to be stored during the computation of a layer.
%; hence, we also compute the feature maps' size as the number of filters multiplied by the output shape of a feature map. 
As an example, the size of the input and the output feature maps of the first convolutional layer is $N_{ch} \cdot N_s + N_k \cdot N_s$, which has to be stored contemporaneously in the memory during the computation time.
\new{Note that this memory can be reused for the next layer once the computation of the previous one is completed. Hence, the biggest sum of two consecutive feature maps dictates the required memory footprint.}

We further estimate the complexity of the model as the sum of the number of \gls{macc} operations in the convolutional and fully connected layers. The number of operations in each depthwise convolutional layer is $(k_H \cdot k_W) \cdot (H_{out} \cdot W_{out}) \cdot N_k$,
% \begin{equation}\label{eq:macc_cnn}
%     \hash\text{\gls{macc}} = k \cdot (H_{out} \cdot W_{out}) \cdot n, %n_{in} \cdot n_{out},
% \end{equation}
with $H_{out}$ and $W_{out}$ being respectively the height and the width of the output feature maps. %, $n_{in}$ and $n_{out}$ respectively the number of input and output feature maps. 
While for the separable convolution, we estimate $(H_{out} \cdot W_{out})\cdot ((1 \cdot 16) + N_k) $ \gls{macc} operations.
Finally, the fully connected layer requires $N_{in}\cdot N_{out}$ operations.

\subsection{\gls{eeg} Channel Selection}\label{sec:cs} % and Model Reduction

% This section presents the second main contribution of the paper. Selecting the most relevant \gls{eeg} channels and reducing the number of \gls{eeg} electrodes yields not only more user comfort and less obtrusiveness but also lower power consumption and longer lifetime for a wearable device: fewer circuits, e.g., \glspl{adc}, are used in the acquisition, at the same time the memory footprint and the computational complexity are reduced for the data processing.
% Hence, before deploying the trained models on the \glspl{mcu} for edge inference, we first reduce the number of \gls{eeg} channels by proposing a novel, automatic channel selection method based on the weights of the spatial convolution.
% %to reduce the number of \gls{eeg} channels 
% %while at the same time minimizing the
% %with minimal accuracy loss. 

\new{
We propose an automatic channel selection method based on the spatial filters of \edgeeegnet{} to reduce further the resource requirements.}
%\subsubsection{Automatic Channel Selection Based on Spatial Filters}
%
%We propose a novel method which automatically extracts the most relevant channels for the \gls{mi} task based on the network weights of the proposed architecture.
%
%In the first layer, \edgeeegnet{} performs \mbox{1-D} convolutions along the spatial dimension of the \gls{eeg} channels as illustrated in Fig.~\ref{fig:cnn}. 
%The learned weights of the spatial filters
%, framed with green solid line, 
%are used to select the most relevant electrodes for solving the \gls{mi} task.
%
Given a pretrained model on all $N_{ch}$ \gls{eeg} channels and $W_S$, the weights of the first spatial convolutional layer with dimension $N_{ch}\times N_k$,
%with $D$ the depth multiplier, 
%$n$ being the number of spatial filters, we first compute the L2 vector norms along $n$.
%, i.e. along the dimension corresponding to each \gls{eeg} channel marked by the green rectangular cuboid in Fig.~\ref{fig:cnn}. 
%Hence, 
we compute the $\ell_2$-norm for each \gls{eeg} channel $i_{ch} = 1, 2, ..., N_{ch}$ as

\begin{equation}
    \norm{w_S}_2(i_{ch}) = \sqrt{\textstyle\sum_{n=1}^{N_k}\abs{w_S(i_{ch}, n)}^2},
\end{equation}

%where $\norm{\cdot}_2$ is the L2 vector norm and $\abs{\cdot}$ calculates the absolute value. In case of \eegnet{}, we have $D=2$ depthwise multiplier for $n=8$ filters, i.e. 2 depthwise convolutions with 8 filters each are performed for each input "image" of the temporal feature maps, whereas for \edgeeegnet{} a depthwise convolution with $D=1$ and $n=16$ is applied on the input "image" of the \gls{eeg} signals. Both result in 16 spatial filters.
%
%Now that we obtain $\norm{w_S}_2$ 
which gives the vector's magnitude of the weights corresponding to each \gls{eeg} channel. We subsequently select $\bar{N}_{ch}$ channels with the biggest $\norm{w_S}_2$ amplitudes, with $\bar{N}_{ch}$ being the desired number of \gls{eeg} electrodes. 
Algorithm~\ref{alg:weights_channel_selection} shows the pseudo-code of the proposed method.
%The pseudo-code of the proposed automatic channel selection based on the network weights is explained in Algorithm~\ref{alg:weights_channel_selection}.
%The same applies to BCI Competition IV 2a dataset with 22 \gls{eeg} channels.
%Finally, a new model with reduced number of channels is trained and validated.

\begin{algorithm}[!b]
%\DontPrintSemicolon
\KwData{%input \gls{eeg} signals with $N_{ch}$ channels, 
Pretrained model on all $N_{ch}$ channels.}
\KwResult{%output \gls{eeg} signals with $\bar{N}_{ch}$ desired channels.
Indexes of $\bar{N}_{ch}$ selected channels.}
\Begin{
 Load the pretrained model\;
 Extract $W_S$, the weights of the spatial convolution\;% with size $N_{ch}\times N_k$\;
 %, the weights of the spatial filters, of dimension $(N_{ch}, D, n)$ with $D$ the depth multiplier of the depthwise convolution and $n$ the number of filters per convolution\;
 \For{each \gls{eeg} channel $i_{ch}$ from $0$ to $N_{ch}-1$}{%Initialize $sum$\;
 %\For{each depth multiplier $d$ from $0$ to $D-1$}{
 %\For{each filter $n$ from $0$ to $N_k-1$}
 %{ Calculate the absolute value of $w_S(i_{ch}, d, f)$\; Take the square of the absolute value\; Cumulatively sum up $sum = sum + \abs{w_S(i_{ch}, d, f)}^2$\;}%}
 Calculate the $\ell_2$ norm
 \begin{equation*}
     \norm{w_S}_2(i_{ch}) = \sqrt{\textstyle\sum_{n=1}^{N_k}\abs{w_S(i_{ch}, n)}^2}\;
 \end{equation*}
    }
 Find the indices that sort $\norm{w_S}_2(i_{ch})$ in descending order\;
 Take the first $\bar{N}_{ch}$ indices\;% that index the $\bar{N}_{ch}$ biggest $\norm{w_S}_2(i_{ch})$\;
% Extract the \gls{eeg} signals from the input signals indicated by the $\bar{N}_{ch}$ indices\;
 %\For{each \gls{eeg} channel $i_{ch}$ from $0$ to $N_{ch}-1$}{find the $N$ biggest $\norm{w_S}_2(i_{ch})$}
}
\caption{Pseudo-code for automatic \gls{eeg} channel selection based on spatial filters of the proposed \gls{cnn}.}
\label{alg:weights_channel_selection}
\end{algorithm}

%\subsubsection{Manual Channel Selections} % Baseline Channel Selection % Manual Distributed Selection

\new{For comparison, we implement two additional manual channel selection methods: one based on the whole scalp as in~\cite{Wang2020_memea}; the other one targets the specific brain region of the sensorimotor cortex such that the \gls{eeg} electrodes can be embedded into smart wearable headphones, e.g., Versus headset~\cite{Wyckoff2019versusheadset} or Melomind~\cite{Spinelli2020melomindheadset}, for better user acceptance.} %mbt Smartfones~\cite{mbraintrain}. %The two proposed manual methods are then compared with the automatic approach and the results are discussed in Sec.~\ref{subsec:results_channels}.
%Moreover, we investigate additional model reduction methods to reduce the feature map size of the network.
%
Fig.~1 in the Appendix shows the positions of the 64 electrodes for the \physionetmmmi{} dataset and the 22 channels filled with yellow used in the \bcicompivtwoa{} dataset. %~\ref{fig:channels}
%
%The Physionet EEG \gls{mmmi} Dataset is recorded using the BCI2000 system~\cite{Schalk2004BCI2000:System} according to the international 10-10 system (omitting the channels Nz, F9, F10, FT9, FT10, A1, A2, TP9, TP10, P9, and P10) with a total of 64 electrodes as shown in Fig.~\ref{fig:channels}.
%
%Following the proposition in~\cite{Wang2020_memea}, 
We select three configurations that are evenly distributed over the entire scalp to directly compare with~\cite{Wang2020_memea} on the \physionetmmmi{} dataset:
a) 19 electrodes based on the international 10-20 system (excluding A1 and A2), encircled with green dashed lines;
b) 38 electrodes by adding intermediate positions, encircled with solid blue lines;
c) 8 electrodes based on Bitbrain headset, encircled with dash-dotted red lines.
%This allows us to compare directly with~\cite{Wang2020_memea}, which uses the same Physionet dataset. 
The distributed configurations are not evaluated on the \bcicompivtwoa{} dataset because of the reduced number of available electrodes.

%
%First of all, we use the positions in the 10-20 international system except for A1 and A2, yielding a configuration with 19 electrodes encircled with green dashed lines in Fig.~\ref{fig:channels}.
%
%Then the intermediate electrodes are evenly selected yielding the configuration with 38 electrodes encircled with blue solid lines.
%
%Finally, since an increasing number of commercially available \gls{eeg} caps have only 8 electrodes distributed evenly over the whole brain region, we also choose the configuration with only 8 electrodes based on the \gls{eeg} headset by Bitbrain\todo{cite}, encircled with dash-dotted red lines in Fig.~\ref{fig:channels}.

%\subsection{Electrodes Placement over Sensorimotor Cortex} % Electrodes Placement over Sensorimotor Cortex % Manual Headset-based Selection

Furthermore, we manually select the \gls{eeg} channels covering the sensorimotor cortex, where most of the brain's activations during \gls{mm} and \gls{mi} are identified, especially around C3 and C4~\cite{Pfurtscheller1999Event-relatedPrinciples}.
This region is also physically covered by the users' audio headphones that are commonly accepted and worn nowadays for extended time periods. This means we can embed the \gls{eeg} electrodes on a wearable device already widely used in daily life. This setup is much less obtrusive and aesthetically more acceptable than traditional \gls{eeg} caps.
%Moreover, a single wearable device can provide multiple functionalities. 
%The same idea has been previously proposed in literature and implemented in commercial devices for stress relief and concentration training based on neurofeedback~\cite{Wyckoff2019versusheadset, Spinelli2020melomindheadset}. % Wyckoff2015
%To the best of our knowledge, no previous study has investigated channel selection for \gls{mi} task in the view of a \gls{eeg} \gls{mi}-\gls{bmi} headphone setup.
%
Hence, we select subsets of electrodes 
%Based on the observations above, we train and validate our proposed neural network by giving as input subsets of \gls{eeg} channels selected 
from the central line covering the C3 and C4 electrodes, as highlighted in Fig.~1 in the Appendix with the red-shaded background. In addition, we separately and incrementally include subsets of adjacent electrodes towards the front and the back of the head, respectively shaded with the green and blue background color. Table~I in the Appendix reports all the configurations with the selected subsets of channels over the sensorimotor cortex for the \physionetmmmi{} dataset. The same is done for the \bcicompivtwoa{} dataset with the available electrodes. % \ref{tab:headset_channels}

\new{Additional comparisons with \gls{csp}- and Riemannian-based methods are discussed in Sec.~\ref{sec:results}.}

%and  \todo{mental state, focus, stress, neurofeedback, meditation, concentration, training, productivity and mental health (particularly ADHD treatment), Sleep better, have better focus and improve your problem solving, Anti-stress Better sleep Sharper focus} 

% \subsection{Additional Model Reduction}

% Besides the channel selection methods described above, we also explore two additional approaches to reduce the input feature map size. 

\subsection{Embedded Implementation}\label{sec:embedded_impl}
%This section summarizes the third main contribution of the paper.
%We proceed with the deployment on low-power \glspl{mcu} for enabling long-term smart wearables for \gls{mi}-\glspl{bci}. 
%As argued in Sec.~\ref{sec:}
% related  work  on  low power  MCU,  deployment,  and quantization
%quantization has become an increasingly common practice nowadays to deploy neural networks for edge computing. The growing popularity is due to the fact that quantized weights and/or activations reduce the memory footprint required for the deployment and it is possible to exploit vectorial instructions of the underlying microprocessor's architecture to increase the energy efficiency. For example, with 8-bit quantization, the memory requirement is reduced by a factor of 4$\times$ compared to 32-bit representations and the 4-way \gls{simd} instructions can speed up the computation by an ideal factor of 4$\times$. 
%
\new{To optimally deploy \edgeeegnet{} on low-power \glspl{mcu}, we first quantize the models to 8-bit representations using quantization-aware training to reduce further the memory footprint and to exploit the hardware extensions. Subsequently, we deploy the quantized networks on the selected \glspl{mcu} and measure the inference runtime and the power consumption.}
%by exploiting its hardware architecture, i.e., \gls{isa} extensions for digital signal processing and the parallel compute engine. %We finally compare our implementation with commercially available solutions.

\subsubsection{Quantization}

We quantize both weights and activations of \edgeeegnet{}, including the input signals, from 32-bit floating-point to 8-bit fixed-point values.
\new{To preserve the classification accuracy, we perform quantization-aware training.}
The addition of the quantization layers is shown in Fig.~\ref{fig:cnn}. %, and the softmax activation is removed since it does not affect the predicted output class from the fully connected layer.
%
%First of all, we notice that the non-linearities following the spatial and separable convolutions are \gls{elu} activations. Calculating exponential functions on a low-power microprocessor is computationally more expensive, i.e., more time consuming and more power hungry, than a simple \gls{relu} activation. Therefore, for quantizing and efficiently deploying the network, we modify the original architecture by using \gls{relu} instead of \gls{elu}, as in~\cite{Schneider2020}. 
%
%Table~\ref{tab:quantization_layers} illustrates the adopted modifications and the addition of the quantization layers. 
%Notice that the quantization is introduced only before the convolutional and the FC layers, since the other layers, i.e. the batch normalization, \gls{relu}, and average pooling, can be calculated sequentially right after computing the convolution that precedes them without writing back and reloading entire feature maps back and forth the memory. %Moreover, quantizing these layers would yield an increased quantization error rather than a significant speedup for the overall computation of the network. Thus, treating those activations as 32-bit fixed-point values does not impact the energy-efficiency.
%
The quantization procedure is as follows:
\begin{itemize}
    \item The network is trained in full precision until epoch $t_a$;
    \item At epoch $t_a$, the quantization of the activations starts using the \gls{ste}~\cite{jacob2018quantization}. The network is readjusted on the quantized activations until epoch $t_w$; %i.e., the quantized values are used in the forward pass while the backpropagation is performed with full precision
    \item From epoch $t_w$ until epoch $t_{end}$, the weights are increasingly quantized using \gls{rpr}~\cite{Cavigelli2020RPR:Networks} at a step size of 10.
\end{itemize}

\new{
\subsubsection{Network Deployment}
% Several works~\cite{garofalo2020pulp, Wang_fann-on-mcu} have demonstrated that \gls{pulp}-based \glspl{mcu}, such as GAP8~\cite{gap82018} or Mr. Wolf~\cite{pullini2019wolf}, are at least one order of magnitude more energy-efficient than the popular commercially-available ARM-based microprocessors. In this work, we choose \gls{pulp} Mr. Wolf by virtue of its extreme energy-efficiency, reaching up to 274\,GOp/s/W~\cite{pullini2019wolf}.

We implement the 8-bit quantized networks on RISC-V-based \gls{pulp} Mr. Wolf~\cite{pullini2019wolf} and Vega~\cite{Rossi2021vega}.
%It has been demonstrated to be at least one order of magnitude more energy-efficient than the commercially-available \glspl{mcu}, e.g., ARM-based microprocessors~\cite{Wang2020_memea, Wang_fann-on-mcu}, thanks to its custom \gls{isa} extensions and the 8-core parallel compute cluster. 
The former has been embedded into Biowolf, the \gls{soa} \gls{bmi} system in terms of power consumption, computational capability, and form factor~\cite{Kartsch2019BioWolf:Connectivity_short}. While the latter is a more recent microprocessor with improved design and technology, presenting the \gls{soa} energy efficiency.
They both feature a \gls{soc} domain with a single core for handling peripherals and simple computations and a cluster domain with eight and nine parallel cores for more compute-intensive tasks. 
The cluster cores are based on RV32IMCF \gls{isa} with additional custom extensions for \gls{dsp} applications, including hardware loops, post-incremental load and store instructions, and support for two-way and four-way \gls{simd} operations.
%The compute cluster of Mr. Wolf is equipped with 8 parallel cores with custom \gls{isa} extensions and an L1 \gls{tcdm} of 64\,kB, while an additional 512\,kB of on-chip L2 memory is present in the \gls{soc} domain. A \gls{dma} unit can handle the data transfer between the L1 and the L2 autonomously. Similarly, Vega presents 9 cores in the cluster domain and has a bigger L1 memory of 128\,kB 

%After selecting the \glspl{mcu}, we proceed with embedded deployment of the networks for \gls{mi} classification at the edge.
%
%Previous work~\cite{Schneider2020} has efficiently deployed \eegnet{} on Mr. Wolf. Since the last two layers \textPhi\textsuperscript{3} and \textPhi\textsuperscript{4} of our proposed \edgeeegnet{} are identical to \eegnet{}, the same implementation and optimizations are reused.
%While for the first two layers \textPhi\textsuperscript{1} and \textPhi\textsuperscript{2}, the number of computations and memory usage is significantly reduced in \edgeeegnet{} by switching the temporal and spatial convolutions. Hence, we reimplement the data handling and the computation for the first two layers to efficiently exploit the data locality.
% As in~\cite{Schneider2020}
We propose a manual implementation of the quantized models, as currently, no deployment tools directly support the selected \glspl{mcu}. However, we follow the same layer-by-layer schedule to better demonstrate the deployment feasibility of our models. We \lucar{take inspiration} from~\cite{Schneider2020} and merge the conversion factor of the quantization with the bias and the scaling factor of the batch normalization layers and reorder the batch normalization, the \gls{relu}, and the pooling layers to reduce the number of divisions since they are expensive operations requiring many computational cycles and introduce rounding errors. 
Moreover, we implement a concurrent computation of the feature maps using the parallel cores and exploit the four-way \gls{simd} for executing four \gls{macc} operations on the 8-bit quantized numbers in a single cycle.
%
%We use the optimized \mbox{1-D} cross-correlation functions from the PULP-DSP library~\cite{wang2019dsp} by exporting the network weights in the reversed order \lucar{with respect to convolutions, as they are mathematically equivalent.} 
\lucar{We export the network weights in the reversed order with respect to convolutions to use the optimized \mbox{1-D} cross-correlation functions from the PULP-DSP library~\cite{wang2019dsp}.}
We implement the data handling and the computation of each layer such that the data locality is preserved and the custom \gls{isa} extensions are effectively used.
}
%
%We partially reuse the implementation of~\cite{Schneider2020} with two main modifications and additions. First, we reimplement the data handling and the computation of layers \textPhi\textsuperscript{1} and \textPhi\textsuperscript{2} to efficiently exploit the data locality and use the \gls{simd} operations. Second, we further reduce the number of divisions by replacing them with multiplication and bit-shift operations, since the latter are more energy-efficient. 
%Different from~\cite{Schneider2020}, we reimplement the data handling and the computation of layers \textPhi\textsuperscript{1} and \textPhi\textsuperscript{2} to efficiently exploit the data locality and use the \gls{simd} operations.
% More specifically,
% \begin{align}
%     \tilde{y}=\frac{\tilde{x}\cdot \tilde{w} + B}{F} & = (\tilde{x} \cdot \tilde{w} \cdot \frac{2^S}{F} + \frac{B\cdot 2^S}{F})\gg S \\
%     & = (\tilde{x} \cdot \tilde{w} \cdot \tilde{F} + \tilde{B})\gg S
% \end{align}
% with $\tilde{x}$, $\tilde{w}$, and $\tilde{y}$ being respectively the 8-bit quantized input, weight, and output values, $B$ and $F$ the bias and the scaling factor, and $S$ the bit-shift value.

%We deploy the models with full-channel configuration for all three tasks and the ones with the lowest number of channels and the highest accuracy for the 2-class task to demonstrate the advantage of channel selection in reducing the energy consumption with experimental measurements.

%We measure the power consumption and the inference runtime of the deployed models using the Keysight N6705B power analyzer.

\section{Experiments and Results}\label{sec:results}

\new{
%This section presents the experimental protocols and the obtained results. 
We use Keras with TensorFlow v1.11 backend to train full-precision models. We apply our methods to two \gls{soa} datasets to tackle both inter-session and inter-subject variability. 
%More details on the datasets can be found in the Appendix.
}

%\section{Datasets}\label{sec:dataset}

% \gls{eeg}-based \glspl{bmi} suffer from inter-session and inter-subject variability that poses considerable challenges to designing accurate models.
% %
%We use two \gls{soa} \gls{mi} datasets to tackle both inter-session and inter-subject variability. More details on the datasets can be found in the Appendix. %, respectively the IV-2a from the famous BCI Competition~\cite{Brunner2008BCIA,tangermann2012_bcicompreview} and the newer and much larger Physionet \gls{eeg} \gls{mmmi}~\cite{Schalk2004BCI2000:System,goldberger2000physiobank}. 

\paragraph{BCI Competition IV-2a~\cite{Brunner2008BCIA,tangermann2012_bcicompreview}}

Four \gls{mi} tasks are recorded from 9 subjects 
using 22 \gls{eeg} channels, 
namely the imagined movements of the left hand (L), right hand (R), both feet (F), and tongue (T).
%22 \gls{eeg} electrodes are used, as highlighted in yellow in Fig.~\ref{fig:channels}. 
\new{The data is collected on two days (i.e., sessions) at 250\,Hz sampling frequency and provided after a bandpass filter between 0.5\,Hz and 100\,Hz.
Each recording session contains 288 trials, of which 9.41\% are marked by an expert as artifacts and are removed.}
We perform 2-, 3-, and 4-class classification using L/R, L/R/F, and L/R/F/T \gls{mi} tasks, respectively, and consider a time window of 3\,s starting from the appearance of the \gls{mi} cue.
%More details are described in the Appendix.
\new{We conduct subject-specific validation using the \gls{eeg} data of the first session for training and the second for testing, as stated in the competition rules. % to account for inter-session variability
The models are trained with the Adam optimizer for 500 epochs with a batch size of 32, a fixed learning rate of 0.001, and 1e-7 as the default epsilon value for numerical stability. The cross-entropy loss is used.
We repeat the experiments 25 times and report the average results to consider the training variability. We calculate both classification accuracy and Cohen's kappa score as it is often done with this dataset.}

\paragraph{Physionet EEG~\gls{mmmi}~\cite{Schalk2004BCI2000:System,goldberger2000physiobank}}
The data is recorded at 160\,Hz using 64 electrodes. % according to the international 10-10 system. 
105 subjects are effectively considered from this dataset. 
Every subject performed a total of six \gls{mi} runs: three runs of the left fist (L) against the right fist (R) and three runs of both fists (B) against both feet (F). We additionally extract windows of 3\,s from the baseline runs with eyes open to obtain trials with resting \gls{eeg} data (0).
\new{In total, we have 21 trials per class per subject.}
As in~\cite{Dose2018AnBCIs}, we perform 2-, 3-, and 4-class \gls{mi} classification using L/R, L/R/0, and L/R/0/F \gls{mi} tasks, respectively, and consider a time window of 3\,s starting from the appearance of the cue.
%For more details please refer to the Appendix.
\new{We use 5-fold \gls{cv} across the subjects, i.e., the model is trained on the data from a subset of subjects and is validated on the data from the remaining unseen subjects. %, \lucar{and this procedure is repeated five times on five non-overlapping subsets.}
The models are trained with Adam optimizer for 100 epochs with a batch size of 16. A fixed learning rate scheduler reaches the best performance; more precisely, the learning rate is set to 0.01, 0.001, and 0.0001 at epochs 0, 40, and 80, respectively.
Cross-entropy loss is used.
\lucar{Similar to the \bcicompivtwoa{} dataset, we repeat the 5-fold \gls{cv} procedure five times and obtain an average accuracy over 25 experiments to account for the training variability.} % Hence, we repeat the 5-fold \gls{cv} procedure 5 times and report the average accuracy.}
}

%In this section, we present and discuss the performance of the proposed energy-efficient \gls{cnn} in terms of classification accuracy, computational complexity, and memory footprint. Moreover, the outcome of the novel channel selection method is presented and compared to manual baselines and related works. Finally, the quantization results are reported, followed by experimental measurements of the power consumption and the runtime of the deployed models on the selected edge platform.
%
%The classification accuracy is reported as an average over the repetitions and across the 9 subjects for the \bcicompivtwoa{} and the 5 \gls{cv} folds for the \physionetmmmi{} dataset. The Cohen's kappa score is also calculated for the \bcicompivtwoa{} as it is often done with this dataset.

\begin{table}[b]
\setlength{\tabcolsep}{2.5pt}
\caption{\new{Resource estimation considering all EEG channels and the 4-class task. The two biggest consecutive feature maps are marked in italic and the total resource requirements in bold.}}
%BCI Comp. IV-2a: $N_{ch}$=22, $N_s$=750, $N_k$=32, $N_f$=64. Physionet \gls{mmmi}: $N_{ch}$=64, $N_s$=480, $N_k$=16, $N_f$=128.
\label{tab:netarch:resources}
\begin{tabular}{@{}lrrrrrr@{}}
\toprule
& \multicolumn{3}{c}{BCI Comp. IV-2a}   & \multicolumn{3}{c}{Physionet \gls{mmmi}} \\ 
& \multicolumn{3}{c}{\scriptsize $N_{ch}$=22,\,$N_s$=750,\,$N_k$=32,\,$N_f$=64} & \multicolumn{3}{c}{\scriptsize $N_{ch}$=64,\,$N_s$=480,\,$N_k$=16,\,$N_f$=128} \\ 
%& \multicolumn{3}{c}{\begin{tabular}[c]{@{}r@{}}$N_{ch}$=22, $N_s$=750,\\ $N_k$=32, $N_f$=64\end{tabular}} & \multicolumn{3}{c}{\begin{tabular}[c]{@{}r@{}}$N_{ch}$=64, $N_s$=480,\\ $N_k$=16, $N_f$=128\end{tabular}} \\ 
\cmidrule(l){2-4} \cmidrule(l){5-7}
%Layer & \#params & \begin{tabular}[c]{@{}r@{}}f. map \\size\end{tabular} & \#\acrshort{macc} & \#params & \begin{tabular}[c]{@{}r@{}}f. map \\size\end{tabular} & \#\acrshort{macc} \\
Layer & \#params & \#features & \#\acrshort{macc} & \#params & \#features & \#\acrshort{macc} \\
%\cmidrule(){1-1} \cmidrule(l){2-4} \cmidrule(l){5-7}
\midrule
Input & & \textit{16,500} & & & \textit{30,720} &  \\
\textPhi\textsuperscript{1} & 704+128  & \textit{24,000}     & 528,000                        & 1,024+64 & \textit{7,680}      & 491,520                        \\
\textPhi\textsuperscript{2} & 2,048+128 & 3,000      & 1,536,000                      & 2,048+64  & 960        & 983,040                        \\
\textPhi\textsuperscript{3} & 1,536+128   & 176        & 96,000                         & 512+64   & 112        & 30,720                         \\
\textPhi\textsuperscript{4} & 1,412      & 4          & 704                            & 484      & 4          & 448                            \\
\textbf{Tot.} &  \textbf{6,084}    & \textbf{40,500}     & \textbf{2,209,408}                      & \textbf{4,228}    & \textbf{38,400}     & \textbf{1,505,728}                      \\ 
\bottomrule
\end{tabular}
\end{table}

\begin{table}[t]
\setlength{\tabcolsep}{2.5pt}
\caption{\new{Comparison of accuracy and kappa values ($\kappa$) in full-channel, full-precision configuration. The publication year follows the apostrophe. The total number of parameters, the maximum number of consecutive features, and the number of \gls{macc} operations are estimated on the 4-class task when available, otherwise on the 2-class task. The proposed solution is highlighted in bold.}} % (e.g., '22 stands for year 2022)
\centering
\label{tab:background}
{
  \begin{threeparttable}
\begin{tabular}{@{}lrrrrr@{}}
\toprule
 & \multicolumn{2}{c}{Accuracy [\%]\,/\,$\kappa$} & \multirow{2}[3]{*}{\begin{tabular}[c]{@{}r@{}}Tot. \#\\ params\end{tabular}} & \multirow{2}[3]{*}{\begin{tabular}[c]{@{}r@{}}Max. \#\\ cons. f.\end{tabular}} & \multirow{2}[3]{*}{\begin{tabular}[c]{@{}r@{}}Tot. \#\\ MACC\end{tabular}} \\ 
\cmidrule(lr){2-3}
  & \multicolumn{1}{r}{2-}   & \multicolumn{1}{r}{4-class} &  &  & \\ 
\midrule
BCI Comp. IV-2a \\
\hspace{1mm} CSP'15\,\cite{das2015smcconf} & 80.56\,/\,-  & -\,/\,- & - & - \\ %CH SEL
\hspace{1mm} MRC'18\,\cite{Hersche2018FastFeatures} & -\,/\,- & 75.47\,/\,- & - & - & - \\ % tot params 32.6\,k
\hspace{1mm} \acrshort{fbcsp}'20\,\cite{das2020fbcspovr} & -\,/\,- & -\,/\,0.65 & - & - & - \\
\hspace{1mm} \new{JSTFD+LDA'20\,\cite{Jiang2020_csp}} & 79.6\,/\,- & -\,/\,- & - & - & - \\
\hspace{1mm} \new{MI-TS'20\,\cite{Wang2020_mi-ts}} & 86.01\,/\,- & -\,/\,0.61 & - & - & - \\
\hspace{1mm} \new{MRC'21\,\cite{Wang2021mrc_iscas,Wang2021mrc_tbiocas}$^\dagger$} & -\,/\,- & 75.10\,/\,- & - & - & - \\ % tot params 4.55\,k
\hspace{1mm} S. ConvNet'17\,\cite{Schirrmeister2017DeepVisualization}$^{||}$ & 79.90\,/\,- & 74.31\,/\,0.66 & 47.3\,k & 1,013\,k & 63.0\,M \\
\hspace{1mm} \eegnet{}'18~\cite{Lawhern2018EEGNet:Interfaces}$^{||}$ & -\,/\,- & 71.30\,/\,- & 2.63\,k & 223\,k & 12.98\,M \\
\hspace{1mm} MSFBCNN'19\,\cite{wu2019_MSFBCNN} & -\,/\,- & 75.80\,/\,- & 155\,k & 5,775\,k & 202\,M \\
\hspace{1mm} \new{MCNN'19\,\cite{Amin2019_MCNN}} & -\,/\,- & 75.7\,/\,- & 14\,M & 574\,k  & 103\,M \\
\hspace{1mm} EEG-TCNet'20\,\cite{ingolfsson2020eegtcnet}$^\ddagger$ & -\,/\,- & 77.35\,/\,0.70 & 4.27\,k & 396\,k & 6.8\,M \\
\hspace{1mm} FB3DCNN'21\,\cite{bang2021fb3dcnn} & 86.96\,/\,- & -\,/\,- & 46\,M & 472.1\,k & 62.3\,M \\
\hspace{1mm} \new{EEG-ITNet'22\,\cite{Salami2022_eegitnet}} & -\,/\,- & 76.74\,/\,- & 5.2\,k & 74.3\,k & 7.36\,M \\
%\hspace{1mm} \acrshort{csp}+Riemannian~\cite{chen2020cssimilarity}$^\ddagger$ & -\,/\,- & \textbf{77.82\,/\,0.71} & - & - & Automatic & None & None & - \\ %only CH SEL
%\hspace{1mm} \acrshort{wola}\acrshort{csp}'18\,\cite{belwafi2018_wolacsp} & 78.85\,/\,- & -\,/\,- & - & - & - \\ % 16-bit
%\hspace{1mm} \acrshort{csp}+\acrshort{svm}~\cite{malekmohammadi2019_cspsvm} & 80.55\,/\,- & 67.21\,/\,- & - & - & - & None & 20-bit & \acrshort{fpga} & 11.66\,ms, 83.90\,mW \\ %20-bit
\hspace{1mm} \textbf{\edgeeegnet{}}$^*$  & \textbf{86.32\,/\,0.73} & \textbf{76.03\,/\,0.68} & \textbf{6.08\,k} & \textbf{40.5\,k} & \textbf{2.21\,M} \\
%& \textbf{87.65\,/\,0.75$^\ddagger$} & \textbf{77.18\,/\,0.70$^\ddagger$} & & & & & & & \textbf{10.57\,ms, 9.7\,mW}$^\dagger$ \\
%
Physionet \acrshort{mmmi} \\
\hspace{1mm} S. ConvNet'18\,\cite{Dose2018AnBCIs} & 80.38/\,- & 58.58/\,- & 203\,k & 1,260\,k & 86.12\,M \\
\hspace{1mm} \eegnet{}'20\,\cite{Wang2020_memea}$^\S$ & 82.43\,/\,- & 65.07\,/\,- & 3.2\,k & 277.56\,k  & 31.98\,M \\
\hspace{1mm} CNN'20\,\cite{Tokovarov2020} & {83.26}\,/\,- & -\,/\,- & 235\,k & 122.88\,k & 33.61\,M \\ % 82.34\% with 14 channels
\hspace{1mm} \textbf{\edgeeegnet{}}$^*$ & \textbf{82.79\,/\,-} & \textbf{65.62\,/\,-} & \textbf{4.23\,k} & \textbf{38.40\,k}  & \textbf{1.51\,M} \\
% & 82.51\,/\,-$^{||}$ & 74.21\,/\,-$^{||}$ & 63.93\,/\,-$^{||}$ & & & & & & & \textbf{5.53\,ms, 9.06\,mW}$^\S$ \\
\bottomrule
\end{tabular}
\begin{tablenotes}\footnotesize
\item $^\dagger$ Improved to 76.4\% using different classifiers for each subject.
\item $^\ddagger$ Improved to 83.84\%\,/\,0.78 using subject-specific hyperparameters.
\item $^{||}$ Adapted and reproduced in \cite{bang2021fb3dcnn,ingolfsson2020eegtcnet,Schneider2020}. \hspace{0.2cm} $^\S$ Based on \cite{Lawhern2018EEGNet:Interfaces}.
\item $^*$ Respectively 80.37\%\,/\,0.71 and 74.92\% for 3-class task.
%Reproduced in \cite{bang2021fb3dcnn} for 2-class, \cite{ingolfsson2020eegtcnet} for 4-class, \cite{Schneider2020,Wang2020_memea}, respectively.
%\item $^*$ Reproduced in \cite{Schneider2020,Wang2020_memea}.
%\hspace{0.3cm} $^\ddagger$ Mean accuracy with subject-specific number of selected channels. 
%\item $^\dagger$ Subject 8 with 6 \gls{eeg} channels for the 2-class task. \hspace{0.2cm} $^{||}$ Respectively with 10, 20, and 18 selected channels. \hspace{0.3cm} $^\S$ 10 channels for the 2-class task.
\end{tablenotes}
\end{threeparttable}
}
\end{table}

\subsection{Network Architecture}

Based on the performance achieved with hyperparameters tuning, we choose the filter size $N_f$ to be 64 and 128 and the number of filters $N_k$ to be 32 and 16 for the \bcicompivtwoa{} and the \physionetmmmi{} dataset, respectively. 
%
%Table~\ref{tab:cnn} demonstrates the number of parameters, the feature maps size, and the estimated number of \glspl{macc} for each convolution block of \edgeeegnet{}, computed following the methods presented in Sec.~\ref{sec:net_archi}.
As reported in Table~\ref{tab:netarch:resources}, our model requires around 6.1\,k and 4.2\,k parameters in full-channel configuration. 
%The respective total feature map size is 43.68\,k and 39.476\,k, with the input layer and the first layer being the most demanding ones. 
%Adopting the most common layer-by-layer computation, we need to store the input and the output feature maps of a layer during inference execution. Hence, the two biggest consecutive feature maps dictate the required memory footprint. In our case, the execution of \edgeeegnet{} requires up to 40.5\,k and 38.4\,k features to be stored, respectively for the \bcicompivtwoa{} and the \physionetmmmi{} dataset.
\new{The two biggest consecutive feature maps contain 40.5\,k and 38.4\,k feature values for the \bcicompivtwoa{} and the \physionetmmmi{} datasets.
This means that the execution of \edgeeegnet{} requires the memory storage for only 46.6\,k and 42.6\,k values, comprehensive of both model parameters and the maximum number of features during inference.
%It is at least one order of magnitude less than most related works reported in Table~\ref{sec:related}.
\lucar{It is at least 1.7$\times$~\cite{Salami2022_eegitnet} and up to two orders of magnitude~\cite{wu2019_MSFBCNN,Amin2019_MCNN,bang2021fb3dcnn} less than the related works reported in Table~\ref{sec:related}.}
%
%The accuracy values are 86.32\%, 80.37\%, 76.03\% (\bcicompivtwoa{}), and 82.79\%, 74.92\%, 65.62\% (\physionetmmmi{}), for 2-, 3-, 4-class tasks, respectively.
}

{When comparing the computational complexity, \edgeeegnet{} requires the least number of \gls{macc} operations,} i.e., down to 2.21 million for the \bcicompivtwoa{} dataset and 1.51 million for the \physionetmmmi{} dataset. In contrast to the most compute-intensive related works~\cite{wu2019_MSFBCNN,Dose2018AnBCIs}, our models require \textit{91$\times$ and 57$\times$ fewer computations while being 0.23\% and 7.04\% more accurate} on the \bcicompivtwoa{} and \physionetmmmi{} datasets, respectively.
\new{{More recent works have reduced the computational complexity~\cite{ingolfsson2020eegtcnet,bang2021fb3dcnn,Salami2022_eegitnet};} however, our model still requires \textit{at least 3$\times$ fewer computations}. This directly correlates with the computational latency during inference time, as shown in Sec.~\ref{sec:res:embedded}. %This will be reflected in the inference time discussed later in Sec.~\ref{sec:res:embedded}.
}

\new{Despite the significantly fewer resource requirements, the accuracy of \edgeeegnet{} is overall comparable to recent \gls{soa} models. More specifically, the accuracy values are 86.32\%, 80.37\%, 76.03\% (\bcicompivtwoa{}), and 82.79\%, 74.92\%, 65.62\% (\physionetmmmi{}), respectively for 2-, 3-, 4-class tasks. 
{The highest 4-class accuracy on the \bcicompivtwoa{} dataset following the competition rules is around 77\% without additional subject-specific hyperparameters tunings~\cite{ingolfsson2020eegtcnet}. %With more tricks, the accuracy has been increased to around 84\%~\cite{ingolfsson2020eegtcnet}. 
%However, this does not diminish our contribution as these tricks can be easily applied to our network while still being the least resource-demanding. As our main goal is to present a ultra-low-power embedded \gls{bmi} solution, we leave additional tricks to future investigations. 
%More tricks have been used to further increase this accuracy, of which details we discuss in Sec.~\ref{sec:discussion}.
The authors have improved the accuracy to about 84\% with variable networks. 
As more details are needed to understand the reason behind this accuracy increase, we discuss this comparison separately in Sec.~\ref{sec:discussion}. 
More recent works achieve 86.96\% and 76.74\% 2- and 4-class accuracy values (less than 0.71\% difference from ours) but demand significantly more resources, \lucar{i.e., 997.3$\times$ and 1.7$\times$ more parameters and features, and 28.2$\times$ and 3.3$\times$ more \glspl{macc}~\cite{bang2021fb3dcnn,Salami2022_eegitnet}.}
%Note that all these networks are impossible to be deployed using the standard layer-by-layer computation on the ultra-low-power Mr. Wolf \gls{mcu}, as discussed later in Sec.~\ref{sec:res:embedded}. % this sentence not anymore valid with eeg-itnet
}
%The accuracy is further improved when the automatic channel selection is applied, as reported in the next subsection.
}

\subsection{\gls{eeg} Channel Selection}\label{subsec:results_channels}

\input{02_figures/F_cs_acc_plots}
\begin{table*}[t]
\setlength{\tabcolsep}{2.6pt}
\caption{\new{Classification accuracy (\%)\,/\,kappa score with all channels and the best accuracy\,/\,kappa score for each subject with the corresponding number of selected channels (in brackets) on the IV-2a dataset and comparison with related works. The best accuracy values are highlighted in bold. The publication year is indicated with the apostrophe.}}
%\caption{Classification accuracy (\%)\,/\,kappa value for each subject of BCI Competition IV-2a dataset with all channels and best number of channels and comparison with related works.}
\centering
\label{tab:bciRes}
{
\footnotesize
%\begin{tabular}{@{}lr@{\hspace{\tabcolsep}}rr@{\hspace{\tabcolsep}}rr@{\hspace{\tabcolsep}}rr@{\hspace{\tabcolsep}}rr@{}}
\begin{threeparttable}
\begin{tabular}{@{}lrrrrrrrrrr@{}}
\toprule
        & \multicolumn{5}{c}{2-class}                                                                  & \multicolumn{2}{c}{3-class}                  & \multicolumn{3}{c}{4-class}\\
\cmidrule(lr){2-6} \cmidrule(lr){7-8} \cmidrule(l){9-11} 
& \multicolumn{2}{c}{\textbf{\edgeeegnet{}}} & \multicolumn{2}{c}{CSP'15\,\cite{das2015smcconf}} & \makebox[0pt][r]{\new{M-CSP'19\,\cite{Gaur2019_memdbf}}} & \multicolumn{2}{c}{\textbf{\edgeeegnet{}}} & \multicolumn{2}{c}{\textbf{\edgeeegnet{}}} & \makebox[0pt][r]{CSP+Riem.'20\,\cite{chen2020cssimilarity}} \\
\cmidrule(lr){2-3} \cmidrule(lr){4-5} \cmidrule(lr){6-6} \cmidrule(lr){7-8} \cmidrule(lr){9-10} \cmidrule(l){11-11} 
S. & all ch. & sel. ch. & all ch. & sel. ch. & sel. ch. & all ch. & sel. ch. & all ch. & sel. ch. & sel. ch. \\
%\cmidrule{1-1} \cmidrule(l){2-2} \cmidrule(l){3-3} \cmidrule(l){4-4} \cmidrule(l){5-5} \cmidrule(l){6-6} \cmidrule(l){7-7} \cmidrule(l){8-8} \cmidrule(l){9-9} \cmidrule(l){10-10} \cmidrule(l){11-11} 
\midrule
1       & 84.03\,/\,0.68      & 86.98\,/\,0.74\,(4)        & 90.78\,/\,-               & 83.36\,/\,-\,(9) & 90.78\,/\,-\,(15)       & 89.43\,/\,0.84      & 91.18\,/\,0.87\,(14)       & 83.10\,/\,0.78      & 84.41\,/\,0.80\,(20)       & 87.51\,/\,0.81\,(14)                         \\
2       & 71.15\,/\,0.42      & 72.65\,/\,0.45\,(20)        & 59.85\,/\,-               & 71.83\,/\,-\,(11) & 57.75\,/\,-\,(15)       & 65.74\,/\,0.49      & 68.76\,/\,0.53\,(16)       & 59.27\,/\,0.46      & 60.49\,/\,0.47\,(18)   & 58.32\,/\,0.44\,(18)                      \\
3       & 94.95\,/\,0.90      & 94.95\,/\,0.90\,(22)       & \textbf{97.81\,/\,-}               & \textbf{98.54\,/\,-\,(13)} & 97.08\,/\,-\,(15)       & \textbf{91.67\,/\,0.88}      & \textbf{92.18\,/\,0.88\,(18)}       & \textbf{90.64\,/\,0.88}      & \textbf{90.97\,/\,0.88\,(18)}       & 89.01\,/\,0.86\,(14)                       \\
4       & 74.38\,/\,0.49      & 76.66\,/\,0.53\,(9)        & 68.10\,/\,-               & 74.13\,/\,-\,(3) & 70.69\,/\,-\,(15)        & 75.66\,/\,0.64      & 76.00\,/\,0.64\,(14)       & 69.77\,/\,0.60      & 70.63\,/\,0.61\,(18)       & 71.12\,/\,0.63\,(15)                    \\
5       & 92.00\,/\,0.84      & 93.84\,/\,0.88\,(4)        & 68.88\,/\,-               & 71.11\,/\,-\,(4) & 61.48\,/\,-\,(15)        & 79.17\,/\,0.70      & 80.33\,/\,0.71\,(18)       & 71.83\,/\,0.62      & {73.61\,/\,0.65\,(19)}       & 63.44\,/\,0.49\,(19)                         \\
6       & 79.48\,/\,0.59      & {81.11}\,/\,0.62\,(20)       & 66.67\,/\,-               & 73.14\,/\,-\,(8)  & 70.37\,/\,-\,(15)       & 63.48\,/\,0.45      & 65.56\,/\,0.49\,(20)       & 58.10\,/\,0.44      & 59.91\,/\,0.47\,(11)       & 60.16\,/\,0.53\,(11)                       \\
7       & 90.51\,/\,0.81      & {91.17}\,/\,0.82\,(20)       & 82.14\,/\,-               & 83.57\,/\,-\,(6) & 72.14\,/\,-\,(15)        & 88.49\,/\,0.83      & 88.97\,/\,0.84\,(18)       & 84.71\,/\,0.80      & 85.76\,/\,0.81\,(19)       & \textbf{93.14\,/\,0.92\,(15)}                       \\
8       & \textbf{98.06\,/\,0.96}      & \textbf{98.27\,/\,0.97\,(6)}        & 97.01\,/\,-               & 96.26\,/\,-\,(14) & \textbf{97.76\,/\,-\,(15)}       & 88.18\,/\,0.82      & 88.63\,/\,0.83\,(19)       & 84.55\,/\,0.80 & 86.48\,/\,0.82\,(18)       & 90.43\,/\,0.86\,(17)                        \\
9       & 92.31\,/\,0.85      & 93.26\,/\,0.87\,(14)       & 93.84\,/\,-               & 94.61\,/\,-\,(5) & 94.62\,/\,-\,(15)        & 81.53\,/\,0.72      & 82.05\,/\,0.73\,(20)       & 82.33\,/\,0.76 & 82.36\,/\,0.77\,(18)       & 87.21\,/\,0.83\,(14)                    \\
\cmidrule(lr){2-3} \cmidrule(lr){4-5} \cmidrule(lr){6-6} \cmidrule(lr){7-8} \cmidrule(lr){9-10} \cmidrule(l){11-11} 
Avg.    & 86.32\,/\,0.73      & \textbf{87.65\,/\,0.75\,(13.2)}    & 80.56\,/\,-               & 83.61\,/\,-\,(8.1)  & 79.19\,/\,-\,(15)    & 80.37\,/\,0.71      & \textbf{81.52\,/\,0.72\,(17.4)}    & 76.03\,/\,0.68      & {77.18\,/\,0.70\,(17.7)} & \textbf{77.82\,/\,0.71\,(15.2)} \\ %    & -\,/\,0.57\,(8.22)         & -\,/\,0.66\,(15)      \\
% \multirow{2}{*}{Avg.}    & 86.32\,/\,0.73      & \textbf{87.65\,/\,0.75}    & 80.56\,/\,-               & 83.61\,/\,- & 79.19\,/\,-    & 80.37\,/\,0.71      & \textbf{81.52\,/\,0.72}    & 76.03\,/\,0.68      & 77.18\,/\,0.70   & \textbf{77.82\,/\,0.71} \\
% & & \textbf{(13.22)} & & (8.11) & (15) & & \textbf{(17.44)} & & (17.67) & \textbf{(15.2)} \\
Std.    & 8.96\,/\,0.18        & 8.42\,/\,0.17\,(7.1)      & 14.85\,/\,-                & 11.35\,/\,-\,(4) & 15.85\,/\,-\,(-)     & 9.78\,/\,0.15        & 9.21\,/\,0.14\,(2.2)      & 11.11\,/\,0.15       & 10.86\,/\,0.15\,(2.5)   & 13.52\,/\,0.17\,(2.3) \\ %  & -\,/\,0.15\,(3.29)         & -\,/\,-         \\ 
% \multirow{2}{*}{Std.}    & 8.96\,/\,0.18        & 8.42\,/\,0.17      & 14.85\,/\,-                & 11.35\,/\,- & 15.85\,/\,-    & 9.78\,/\,0.15        & 9.21\,/\,0.14     & 11.11\,/\,0.15       & 10.86\,/\,0.15     & 13.52\,/\,0.17         \\ 
% & & (7.11) & & (3.95) & (-) & & (2.17) & & (2.45) & (2.30) \\
\cmidrule(){1-1} \cmidrule(lr){2-3} \cmidrule(lr){7-8} \cmidrule(lr){9-10}
% \makebox[0pt][l]{\xia{Avg.$^*$}} & & \makebox[0pt][r]{87.24\,/\,0.75\,(8.89)} &  &  &  &  & \makebox[0pt][r]{81.09\,/\,0.72\,(13.67)} & & \makebox[0pt][r]{76.62\,/\,0.69\,(14.22)} & \\
% \makebox[0pt][l]{\xia{Std.$^*$}} & & \makebox[0pt][r]{8.40\,/\,0.17\,(5.97)} &  & &  &  & \makebox[0pt][r]{9.23\,/\,0.14\,(3.89)} & & \makebox[0pt][r]{10.95\,/\,0.15\,(4.26)} & \\
\makebox[0pt][l]{\xia{Avg.$^*$}} & & \makebox[0pt][r]{87.24\,/\,0.75\,(8.9)} &  &  &  &  & \makebox[0pt][r]{81.09\,/\,0.72\,(13.7)} & & \makebox[0pt][r]{76.62\,/\,0.69\,(14.2)} & \\
\makebox[0pt][l]{\xia{Std.$^*$}} & & \makebox[0pt][r]{8.40\,/\,0.17\,(6)} &  & &  &  & \makebox[0pt][r]{9.23\,/\,0.14\,(3.9)} & & \makebox[0pt][r]{10.95\,/\,0.15\,(4.3)} & \\
\bottomrule
\end{tabular}
\begin{tablenotes}\footnotesize
\item \xia{$^*$ Taking the minimum number of channels while allowing a maximum accuracy degradation of 1\%. See Appendix for the subject-specific values.}
\end{tablenotes}
\end{threeparttable}
}
\end{table*}

\begin{table}[!b]
\setlength{\tabcolsep}{2.7pt}
\caption{\new{Comparison of accuracy and resource requirements with channel selection on the \gls{mmmi} dataset.
%Classification accuracy (\%), number of parameters, maximum feature map size, and computational complexity of \edgeeegnet{} and its comparison with related works on the Physionet \gls{mmmi} dataset. 
The estimated resources are calculated for the 4-class task when available, otherwise for the 2-class task. 
%The best solution for each configuration and metric is marked in bold.
The proposed solutions are highlighted in bold.}}
\centering
\label{tab:physioRes}
{
  \begin{threeparttable}
\begin{tabular}{@{}lrrrrrrr@{}}
\toprule
          & \multirow{2}[3]{*}{\#\,ch} & \multirow{2}[3]{*}{\begin{tabular}[c]{@{}c@{}}Tot. \#\\ params\end{tabular}} & \multirow{2}[3]{*}{\begin{tabular}[c]{@{}c@{}}Max. \#\\ cons. f.\end{tabular}} & \multirow{2}[3]{*}{\begin{tabular}[c]{@{}c@{}}Tot. \#\\ MACC\end{tabular}} & \multicolumn{3}{c}{Accuracy}                     \\ \cmidrule(l){6-8} 
          & &  & & & \multicolumn{1}{c}{2-}   & 3-    & 4-class  \\
%\cmidrule{1-5} \cmidrule(l){6-8} 
\midrule
S.ConvNet'18\,\cite{Dose2018AnBCIs} & 64 & 203\,k & 1,260\,k & 86.12\,M & 80.38 & 69.82 & 58.58 \\
CNN'20\,\cite{Tokovarov2020} & 64 & 235\,k & 122.88\,k & 33.61\,M & {83.26} & - & - \\
%Cit.\,\cite{Zhang2020_temporalAttentionGraph} & 64 & - & - & - & 74.71 & - & - \\
EEGNet'20\,\cite{Wang2020_memea}$^\S$     & 64 & {3.17\,k}                                                                & 276.48\,k                                                                          & 48.17\,M                                    %                           & 82.43          & 75.07          & 65.07          \\
                          & 82.29          & 74.46          & 64.85         \\
\textbf{\edgeeegnet{}} & \textbf{64} & \textbf{4.23\,k}   & \textbf{38.40\,k}     & \textbf{1.51\,M}   & \textbf{82.79} & \textbf{74.92} & \textbf{65.62} \\[0.1cm]
%\midrule
EEGNet'20\,\cite{Wang2020_memea}$^\S$ & 38 & 2.76\,k & 164.16\,k & 24.81\,M & 81.86 & 74.12 & 64.65 \\
\textbf{\edgeeegnet{}} & \textbf{38} & \textbf{3.81\,k} & \textbf{25.92\,k} & \textbf{1.31\,M} & \textbf{82.76} & \textbf{74.93} & \textbf{64.91} \\[0.1cm]
CNN'20\,\cite{Tokovarov2020} & 32 & 232\,k & 61.44\,k & 15.82\,M& {82.90} & - & - \\
\textbf{\edgeeegnet{}} & \textbf{32} & \textbf{3.49\,k} & \textbf{24.11\,k} & \textbf{1.26\,M} & \textbf{82.82} & \textbf{74.42} & \textbf{64.38} \\[0.1cm]
EEGNet'20\,\cite{Wang2020_memea}$^\S$ & 19 & 2.35\,k & 82.08\,k & 11.02\,M & 81.95 & 72.41 & 62.55 \\
\textbf{\edgeeegnet{}} & \textbf{19} & \textbf{3.51\,k} & \textbf{16.8\,k} & \textbf{1.16\,M} & \textbf{82.78} & \textbf{74.02} & \textbf{63.69} \\[0.1cm]
%\midrule
% \textbf{\edgeeegnet{}} & 20 & \todo{} & - & - & 82.82 & 74.21 & 63.74 \\
% \textbf{\edgeeegnet{}} & 18 & \todo{} & - & - & 82.84 & 73.94 & 63.93 \\
S.ConvNet'18\,\cite{Dose2018AnBCIs} & 16 & 126\,k & 314.88\,k & 21.6\,M & 78.03 & - & - \\
\textbf{\edgeeegnet{}} & \textbf{16} & \textbf{3.23\,k} & \textbf{15.36\,k} & \textbf{1.14\,M} & \textbf{82.63} & \textbf{73.47} & \textbf{63.47} \\[0.1cm]
%\midrule
S.ConvNet'18\,\cite{Dose2018AnBCIs} & 14 & 123\,k & 275.52\,k & 18.92\,M & 76.66 & - & - \\
CNN'20\,\cite{Tokovarov2020} & 14 & 231\,k & 26.88\,k & 6.68\,M & 82.34 & - & - \\
\new{GCCS'21\,\cite{VARSEHI2021}} & 14 & - & - & - & {83.63} & - & - \\
\textbf{\edgeeegnet{}} & \textbf{14} & \textbf{3.20\,k} & \textbf{14.40\,k} & \textbf{1.12\,M} & \textbf{82.64} & \textbf{72.89} & \textbf{63.30} \\[0.1cm]
%\midrule
\textbf{\edgeeegnet{}} & \textbf{10} & \textbf{3.14\,k} & \textbf{12.48\,k} & \textbf{1.09\,M} & \textbf{82.51} & \textbf{71.55} & \textbf{61.94} \\[0.1cm]
%\midrule
S.ConvNet'18\,\cite{Dose2018AnBCIs} & 9 & 115\,k & 177.12\,k & 12.20\,M & 75.85 & - & - \\
\new{GCCS'21\,\cite{VARSEHI2021}} & 9 & - & - & - & 81.26 & - & - \\
\textbf{\edgeeegnet{}} & \textbf{9} & \textbf{3.12\,k} & \textbf{12.00\,k} & \textbf{1.08\,M} & \textbf{82.06} & \textbf{71.24} & \textbf{61.44} \\[0.1cm]
EEGNet'20\,\cite{Wang2020_memea}$^\S$ & 8 & 2.28\,k & 34.56\,k & 4.30\,M & 78.07 & 68.99 & 58.55 \\
\textbf{\edgeeegnet{}} & \textbf{8} & \textbf{3.33\,k} & \textbf{11.52\,k} & \textbf{1.08\,M} & \textbf{81.60} & \textbf{70.28} & \textbf{60.71} \\[0.1cm]
%\midrule
S.ConvNet'18\,\cite{Dose2018AnBCIs} & 3 & 106\,k & 59.04\,k & 4.13\,M & 73.20 & - & - \\
\textbf{\edgeeegnet{}} & \textbf{3} & \textbf{3.03\,k} & \textbf{9.12\,k} & \textbf{1.04\,M} & \textbf{77.42} & \textbf{61.66} & \textbf{50.12} \\
% \midrule
% \textbf{\edgeeegnet{}} & 64$^\S$ & 4,228 & 39,484 & 1,505\,k & 81.45 & 75.12 & 65.31 \\
% \textbf{\edgeeegnet{}} & 10$^\S$ & \todo{} & - & - &  &  &  \\
\bottomrule
\end{tabular}
\begin{tablenotes}\footnotesize
\item $^\S$ Based on~\cite{Lawhern2018EEGNet:Interfaces}. The reported accuracy is reproduced by averaging over five training/validation repetitions.
%\item $^\ddagger$ Reproduced. Average accuracy over 5 repetitions.
%\item $^\dagger$Network for 4-class task.  
\end{tablenotes}
  \end{threeparttable}
  }
\end{table}

\begin{figure*}
\captionsetup[subfigure]{labelformat=empty}
\centering
\begin{subfigure}{.11\linewidth}
  \centering
  \adjustbox{trim=0.46cm 0.48cm 0.71cm 0.53cm,clip}{%
  \includegraphics[width=1.47\linewidth]{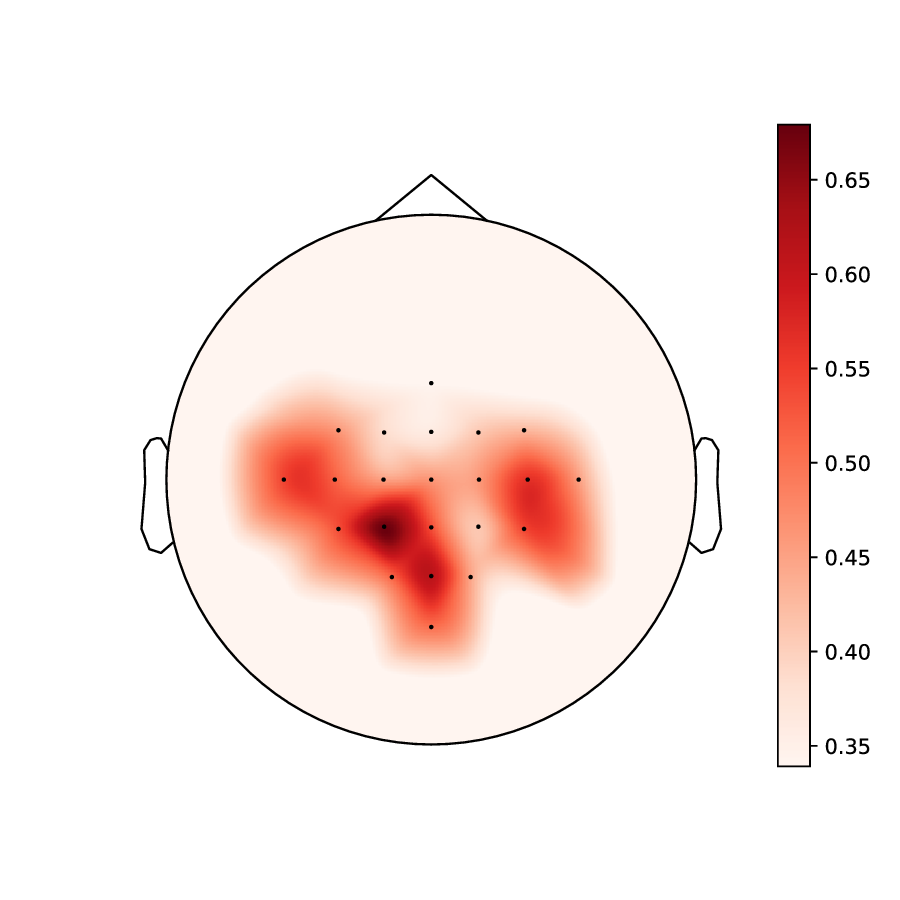}
  }
  \caption{S1: 83.10\%}
\end{subfigure}%
\begin{subfigure}{.11\linewidth}
  \centering
  \adjustbox{trim=0.46cm 0.48cm 0.71cm 0.53cm,clip}{%
  \includegraphics[width=1.47\linewidth]{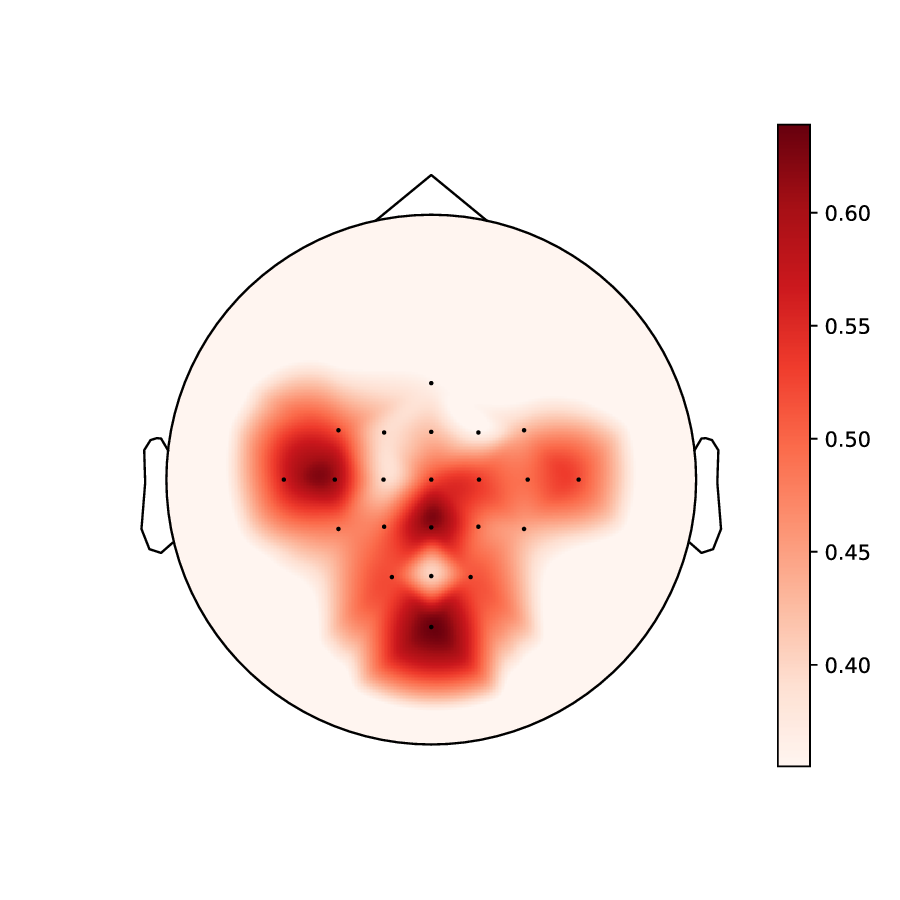}
  }
  \caption{S2: 59.27\%}
\end{subfigure}%
\begin{subfigure}{.11\linewidth}
  \centering
  \adjustbox{trim=0.46cm 0.48cm 0.71cm 0.53cm,clip}{%
  \includegraphics[width=1.47\linewidth]{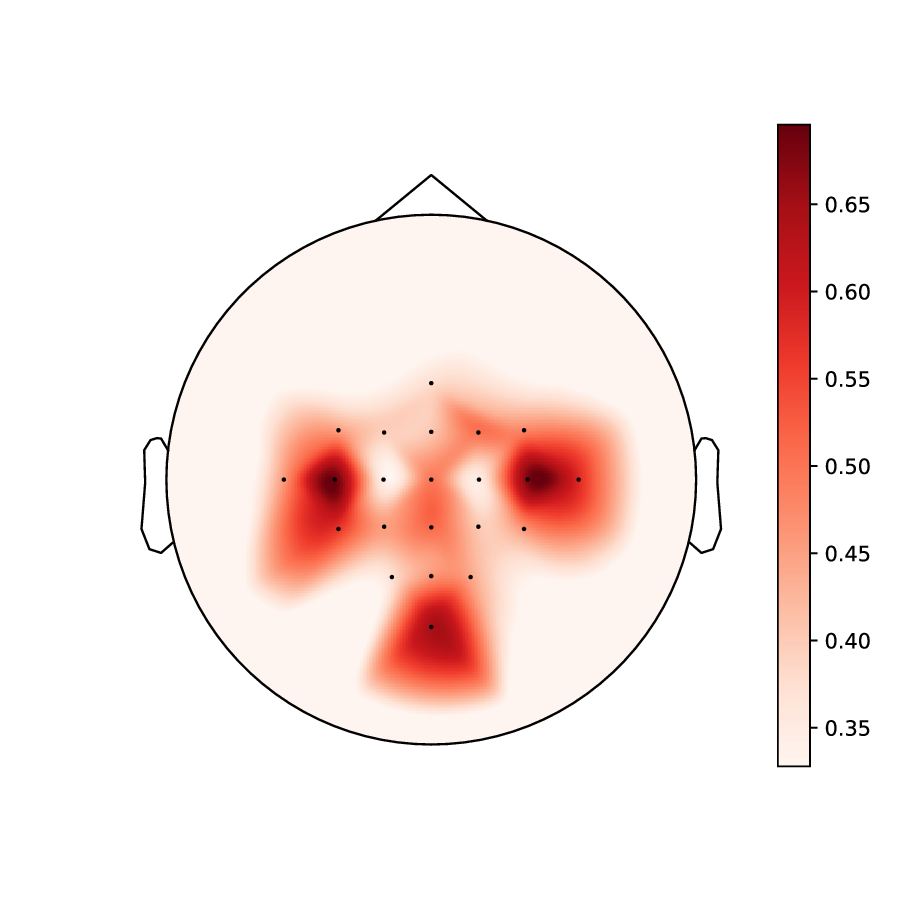}
  }
  \caption{S3: 90.64\%}
\end{subfigure}%
\begin{subfigure}{.11\linewidth}
  \centering
  \adjustbox{trim=0.46cm 0.48cm 0.71cm 0.53cm,clip}{%
  \includegraphics[width=1.47\linewidth]{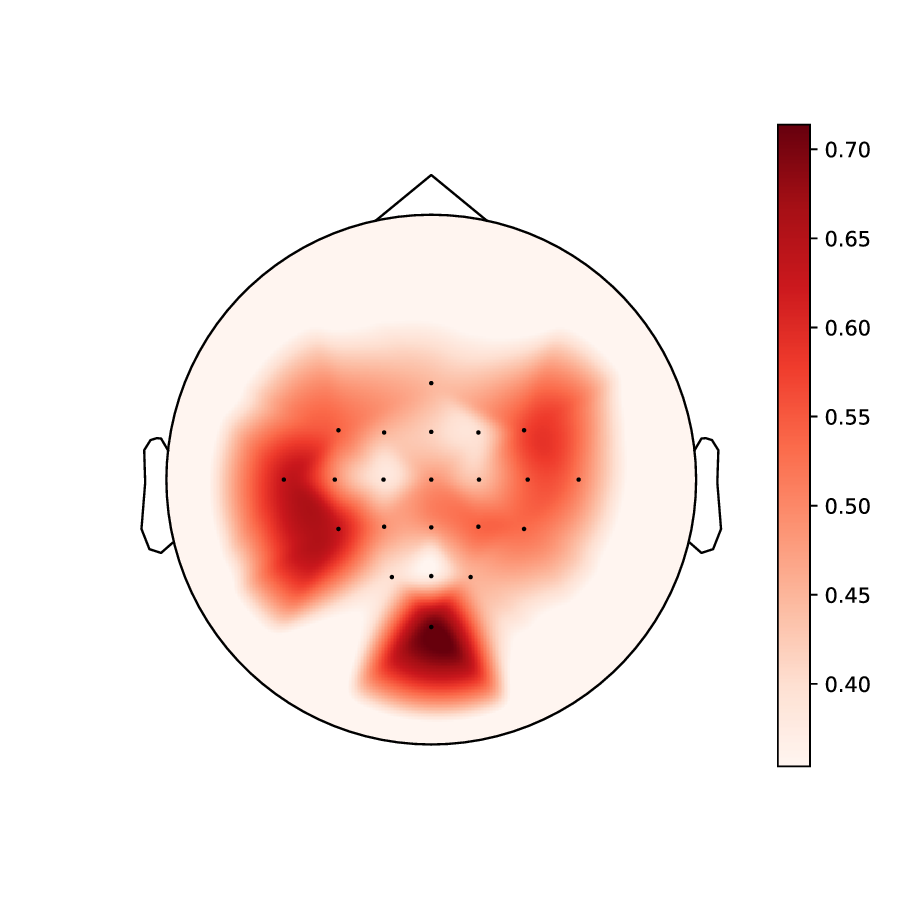}
  }
  \caption{S4: 69.77\%}
\end{subfigure}%
\begin{subfigure}{.11\linewidth}
  \centering
  \adjustbox{trim=0.46cm 0.48cm 0.71cm 0.53cm,clip}{%
  \includegraphics[width=1.47\linewidth]{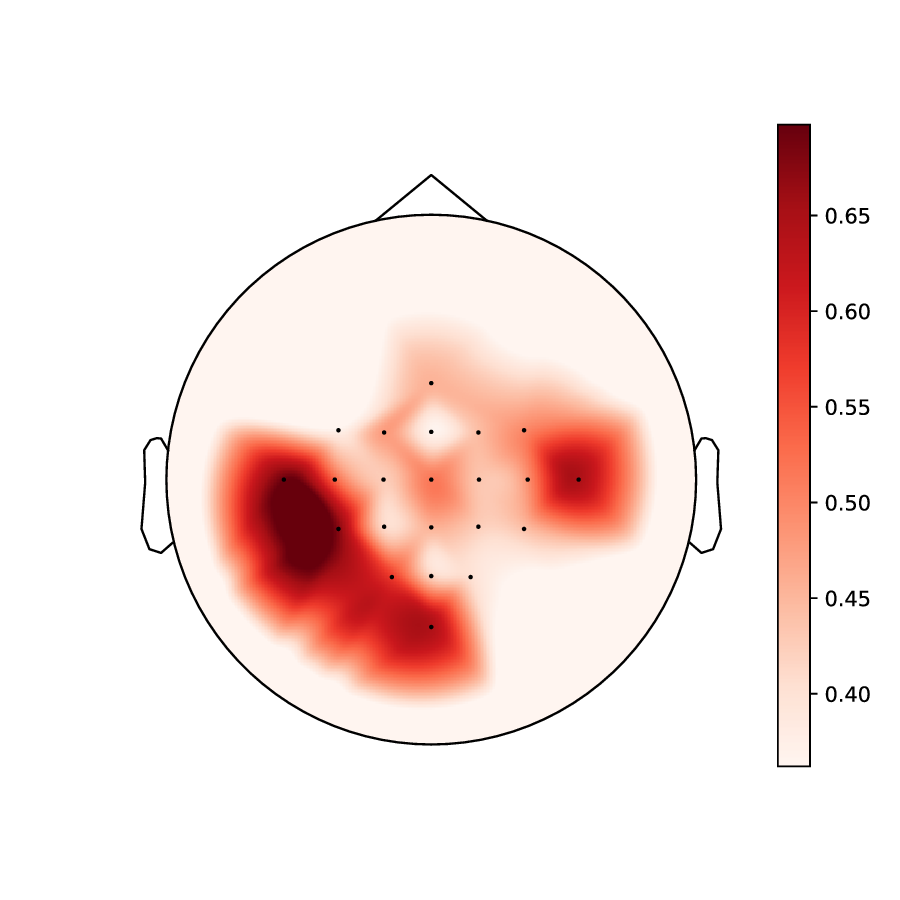}
  }
  \caption{S5: 71.83\%}
\end{subfigure}%
\begin{subfigure}{.11\linewidth}
  \centering
  \adjustbox{trim=0.46cm 0.48cm 0.71cm 0.53cm,clip}{%
  \includegraphics[width=1.47\linewidth]{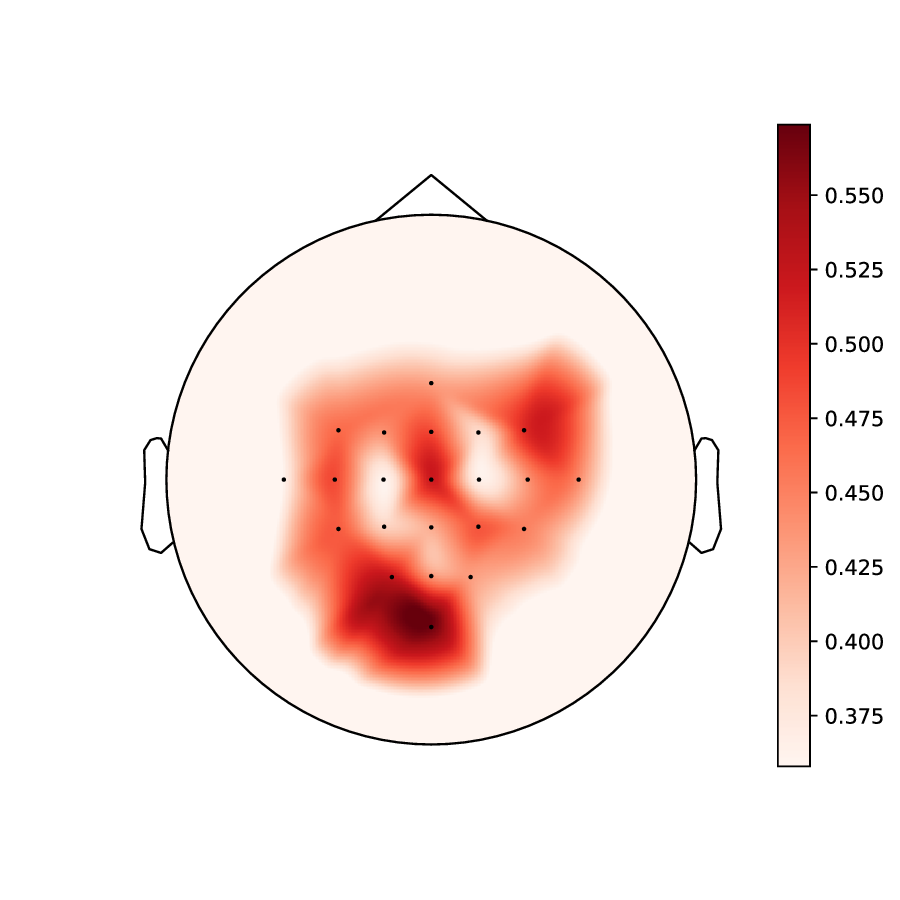}
  }
  \caption{S6: 58.10\%}
\end{subfigure}%
\begin{subfigure}{.11\linewidth}
  \centering
  \adjustbox{trim=0.46cm 0.48cm 0.71cm 0.53cm,clip}{%
  \includegraphics[width=1.47\linewidth]{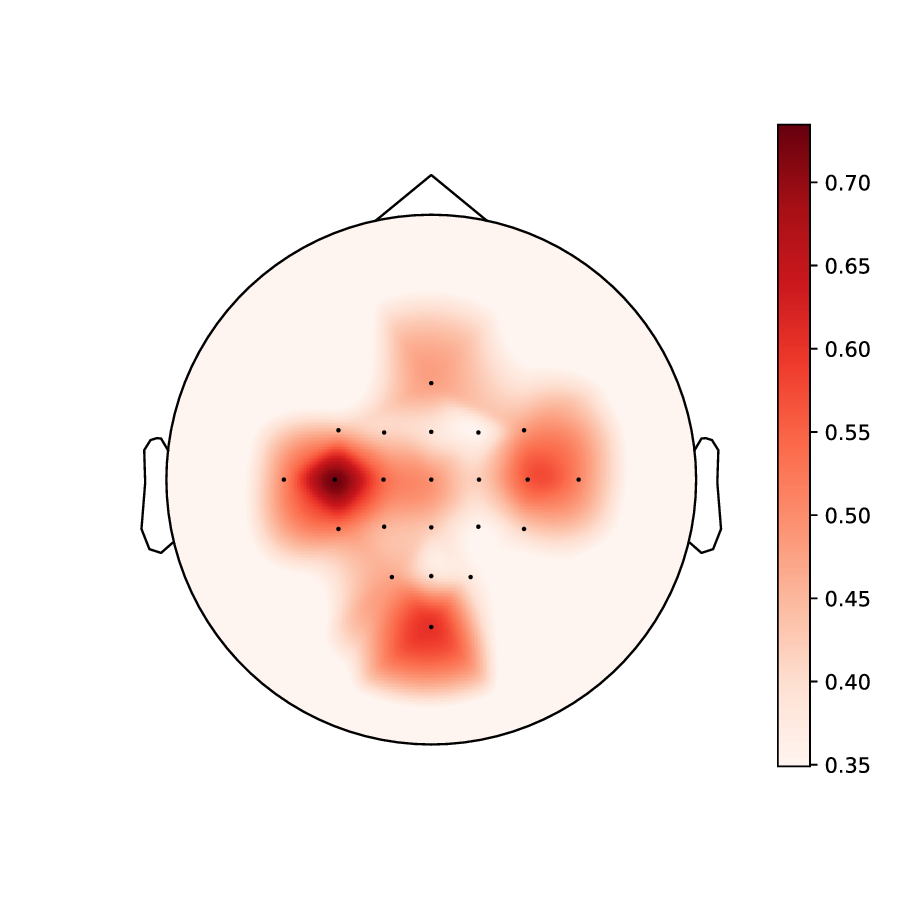}
  }
  \caption{S7: 84.71\%}
\end{subfigure}%
\begin{subfigure}{.11\linewidth}
  \centering
  \adjustbox{trim=0.46cm 0.48cm 0.71cm 0.53cm,clip}{%
  \includegraphics[width=1.47\linewidth]{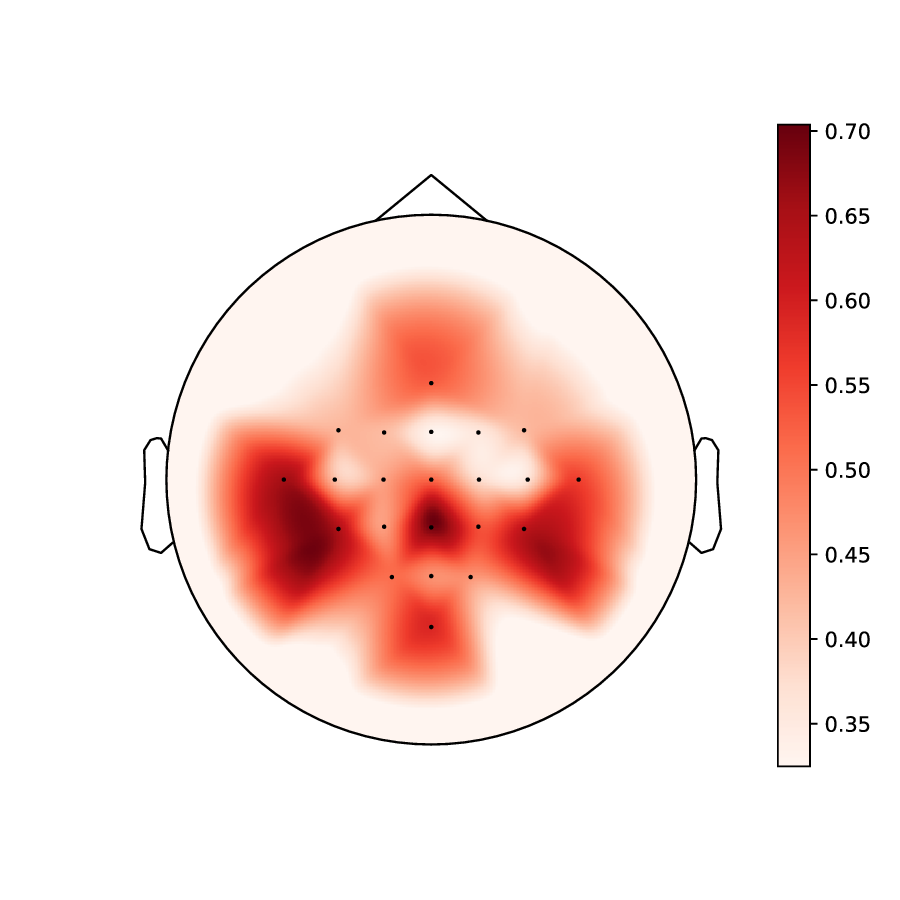}
  }
  \caption{S8: 84.55\%}
\end{subfigure}%
\begin{subfigure}{.11\linewidth}
  \centering
  \adjustbox{trim=0.46cm 0.48cm 0.71cm 0.53cm,clip}{%
  \includegraphics[width=1.47\linewidth]{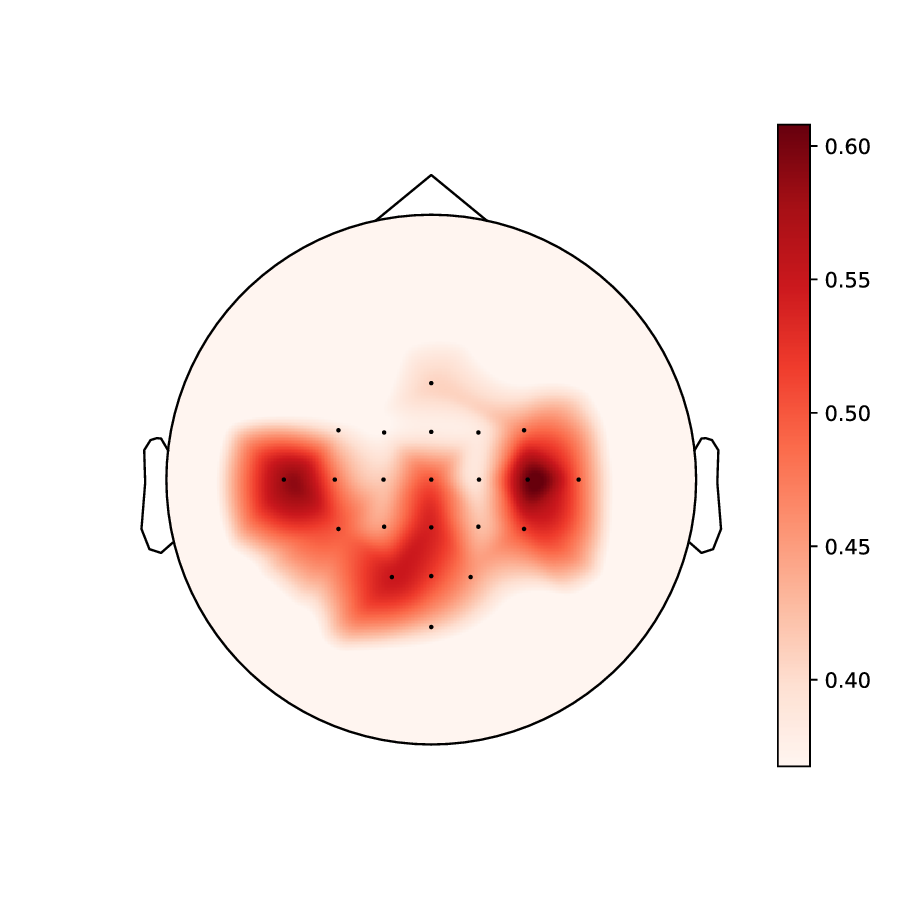}
  }
  \caption{S9: 82.33\%}
\end{subfigure}%
\caption{\new{Heatmaps of $\protect\norm{w_S}_2(i_{ch})$ on the \bcicompivtwoa{} dataset averaged over 25 repetitions with 4-class full-channel accuracy. Darker red color represents higher values, meaning that the networks have learned stronger spatial weights.}} %The average accuracy over nine subjects is 76.03\%.
\label{fig:bci:heatmaps}
\end{figure*}

\begin{figure}
\centering
\begin{subfigure}{.3\columnwidth}
  \centering
  \adjustbox{trim=0.5cm 0.55cm 0.72cm 0.6cm,clip}{%
  %\adjustbox{trim=0.6cm 0.68cm 0.7cm 0.7cm,clip}{%
  %\includesvg[width=1.2\linewidth]{heatmaps/edgeEEGNet_global_class_2_ds1_nch2cs_T3_avg}
  \includegraphics[width=1.2\linewidth]{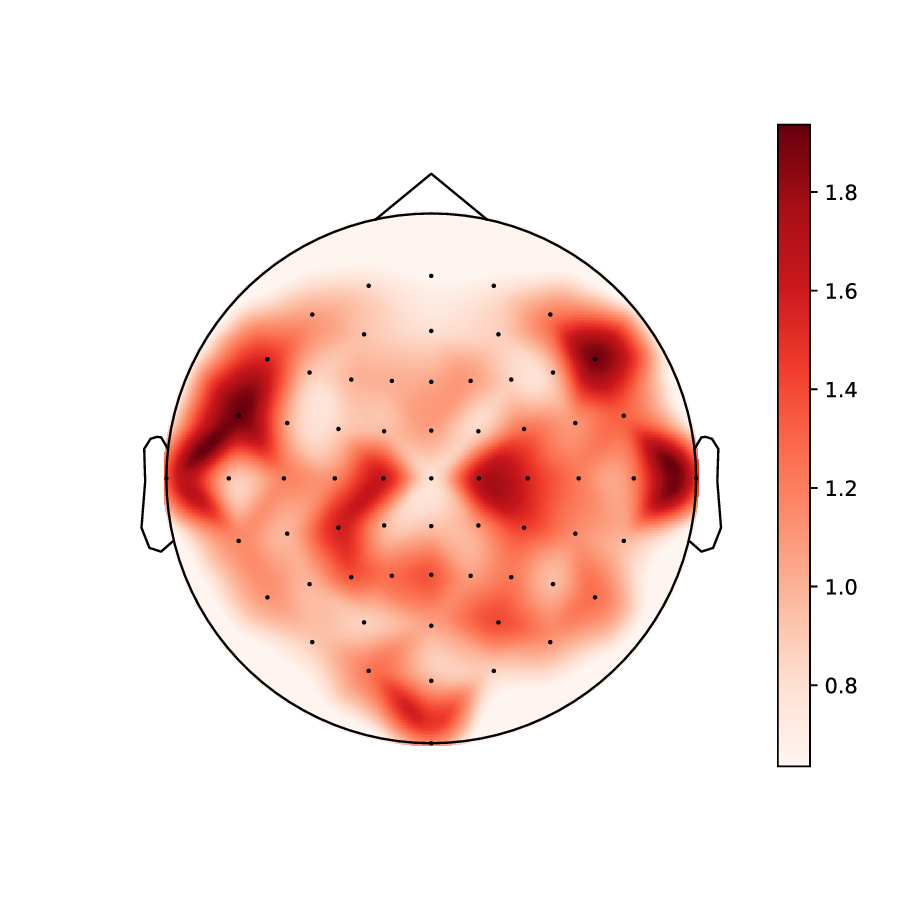}
  }
  \caption{2-class} %: 82.79\%
\end{subfigure}%
\begin{subfigure}{.3\columnwidth}
  \centering
  \adjustbox{trim=0.5cm 0.55cm 0.72cm 0.6cm,clip}{%
  %\adjustbox{trim=0.6cm 0.68cm 0.7cm 0.7cm,clip}{%
  %\includesvg[width=1.2\linewidth]{heatmaps/edgeEEGNet_global_class_3_ds1_nch2cs_T3_avg}
  \includegraphics[width=1.2\linewidth]{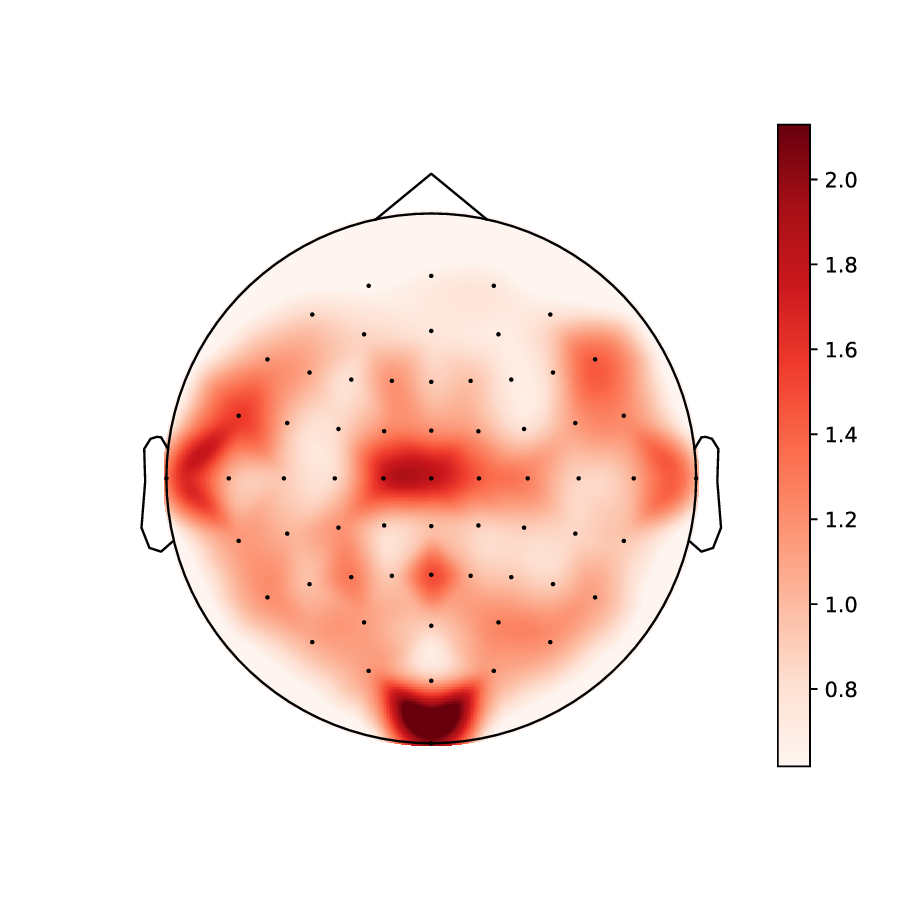}
  }
  \caption{3-class} %: 74.92\%
\end{subfigure}%
\begin{subfigure}{.3\columnwidth}
  \centering
  \adjustbox{trim=0.5cm 0.55cm 0.72cm 0.6cm,clip}{%
  %\adjustbox{trim=0.6cm 0.68cm 0.7cm 0.7cm,clip}{%
  %\def\svgscale{0.5}
  %\includesvg[width=1.2\linewidth]{heatmaps/edgeEEGNet_global_class_4_ds1_nch2cs_T3_avg}
  \includegraphics[width=1.2\linewidth]{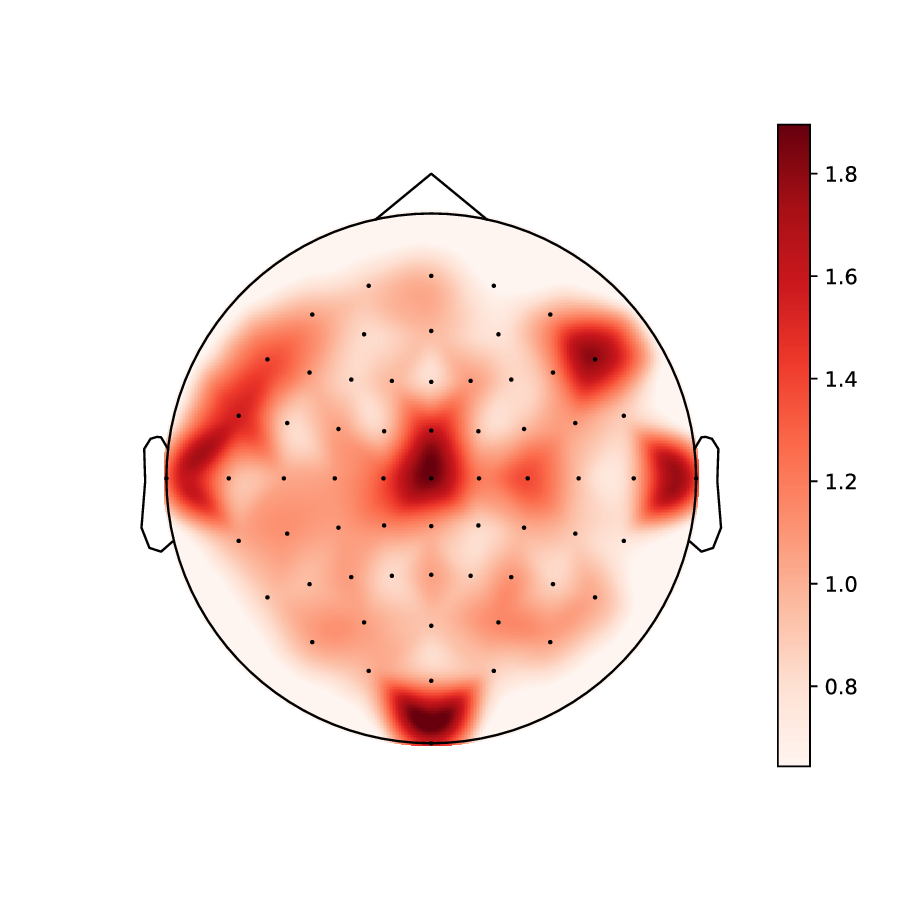}
  }
  \caption{4-class} %: 65.62\%
\end{subfigure}
% \vfill   %%%%%%%%%%%%%%%%%%%%%%%%%%%%%%%%%%%%%%%%%%%%%%%%%%%%%%%%%%%%%%%%%%%
% \begin{subfigure}{.3\columnwidth}
%   \centering
%   %\resizebox{0.5\columnwidth}{!}{%
%   %\adjustbox{trim=0.5cm 0.9cm 1cm 1cm,clip}{%
%   \adjustbox{trim=0.6cm 0.68cm 0.7cm 0.7cm,clip}{%
%   %\def\svgscale{0.5}
%   \includesvg[width=1.5\linewidth]{heatmaps/EEGNet_global_class_4_ds1_nch2cs_T3_avg}
%   }
%   \caption{4 class}
% \end{subfigure}%
% \begin{subfigure}{.3\columnwidth}
%   \centering
%   \adjustbox{trim=0.6cm 0.68cm 0.7cm 0.7cm,clip}{%
%   \includesvg[width=1.5\linewidth]{heatmaps/EEGNet_global_class_3_ds1_nch2cs_T3_avg}
%   }
%   \caption{3 class}
% \end{subfigure}%
% \begin{subfigure}{.3\columnwidth}
%   \centering
%   \adjustbox{trim=0.6cm 0.68cm 0.7cm 0.7cm,clip}{%
%   \includesvg[width=1.5\linewidth]{heatmaps/EEGNet_global_class_2_ds1_nch2cs_T3_avg}
%   }
%   \caption{2 class}
% \end{subfigure}
\caption{\new{Heatmaps of $\protect\norm{w_S}_2(i_{ch})$ on the \gls{mmmi} dataset averaged over 5-fold CV and 5 repetitions. Darker color indicates regions corresponding to stronger learned spatial weights.}}% The first row is with \edgeeegnet{}, the second with \eegnet{}.}
\label{fig:physio:heatmaps}
\end{figure}

As shown in Fig.~\ref{fig:cs:acc}, our automatic method consistently outperforms the manual selection methods. It can reduce, on average, the number of channels down to 9, 11, and 14 across the 9 subjects of the \bcicompivtwoa{} dataset while retaining similar average accuracy (86.21\%, 79.91\%, 75.84\%) as the baseline, represented with solid lines (86.32\%, 80.37\%, 76.03\%), respectively for the \mbox{2-}, \mbox{3-}, and 4-class tasks. In some cases, slightly better accuracy is achieved compared to the full-channel baseline thanks to the subject-specific methodology. 
\new{We report the best accuracy for each subject with the corresponding number of channels in Table~\ref{tab:bciRes}. The average accuracy is improved by up to 1.33\%.}
%In fact, by selecting the best number of channels for each subject, we obtain an improvement in the average accuracy of up to 1.33\%, as reported in Table~\ref{tab:bciRes}. 
%The number of channels can be reduced down to 4, 14, and 11, respectively for \mbox{2-}, \mbox{3-}, and 4-class tasks while improving the accuracy by up to 2.95\% (S1), 1.84\% (S1), and 1.81\% (S6) compared to the full-channel configuration. 
The best performing subjects for \mbox{2-}, \mbox{3-}, and 4-class tasks are S8, S3, and S3, reaching up to 98.27\%, 92.18\%, and 90.97\% with 6, 18, and 18 electrodes, respectively.
\new{
%In general, the simpler the task, the lower the number of selected channels. 
%For example, the 2-class accuracy of S1 and S5 is increased compared to the baseline when only 4 channels are used. Whereas for S3, the best accuracy corresponds to the full-channel configuration, meaning that the trade-off between accuracy and resource usage has to be further assessed for this subject. 
%For S1, S4, S5, and S6, the number of channels can be reduced down to 4, 11, and 14 with increased accuracy. The simpler the task, i.e., 2-class, the lower the number of channels.
Compared to the related works on this dataset, our subject-specific models achieve 4.04\% and 8.46\% better accuracy than~\cite{das2015smcconf} and \cite{Gaur2019_memdbf} for the 2-class task, respectively, and 0.64\% less accuracy than~\cite{chen2020cssimilarity} for the 4-class task.
\xia{The number of selected channels corresponding to the highest accuracy is vastly dependent on the subject, meaning that the trade-off between accuracy and resource usage has to be specifically assessed. An example is reported at the bottom of Table~\ref{tab:bciRes}, where we select the minimum number of channels for each subject while accepting an accuracy degradation of 1\% at most. The average number of channels is further reduced while maintaining similar accuracy values.}}
%Compared to the \gls{soa} with full-channel configuration reported in Table~\ref{tab:background}, we achieve the new \gls{soa} accuracy of 87.65\% and a kappa value of 0.75 for the 2-class task.

%When analyzing the results on the \physionetmmmi{} dataset, we see that 
%the center-front electrodes consistently outperform the center-back ones and achieves slightly better accuracy than using only the central electrodes for 3- and 4-class tasks. This is expected, since the motor cortex is located anterior to the central sulcus in the frontal lobe. 
%the usage of distributed electrodes yields much better results than the headphone configurations, where the electrodes are manually selected over the sensorimotor cortex, especially for the 3- and 4-class tasks. This suggests that regions outside the sensorimotor cortex are also relevant for classifying the 'rest' and 'feet' classes for this dataset. 
%
%On the other hand, 
\new{The results on the \physionetmmmi{} dataset confirm that the automatic method outperforms the manual methods. In general, the simpler the task (i.e., 2-class), the less drop in accuracy. %, represented as horizontal lines in Fig.~\ref{fig:cs:acc:physio}. 
More precisely, we achieve almost the same accuracy as the baseline (82.79\% and 74.92\%) down to 10 and 20 channels (82.51\% and 74.21\%) for the 2- and 3-class classification, respectively.}
%using all the 20 center-front electrodes gives almost the same accuracy as the automated method, which achieves almost the same accuracy as the baseline (82.79\%) down to 10 channels (82.51\%), then the accuracy starts to drop.
For the 4-class task, the accuracy drop is more significant, i.e., \mbox{1.69\%}, when reducing the number of channels to 18.
Table~\ref{tab:physioRes} reports the comparison with the related works. 
%Tokovarov et al.~\cite{Tokovarov2020} use correlation maps from output feature maps of spatial convolution layer to select relevant \gls{eeg} channels. While their full model achieves 0.47\% better accuracy than ours, at the cost of two orders of magnitude more memory requirement and 22$\times$ more complexity, our channel selection method yields less accuracy drop \mbox{(-0.15\%)} than theirs \mbox{(-0.92\%)} using 14 \gls{eeg} channels. The authors did not further reduce the number of channels. %, meaning that with limited number of electrodes, our model achieves 0.3\% better accuracy than~\cite{Tokovarov2020}.
%their channel selection method yields 0.92\% accuracy drop using 14 \gls{eeg} channels, while ours yields only 0.15\% drop using the same number of channels. %, which is 0.3\% better than~\cite{Tokovarov2020}.
\new{The full-channel model by~\cite{Tokovarov2020} achieves 0.47\% better accuracy than ours, at the cost of two orders of magnitude more memory requirement and 22$\times$ more complexity. Whereas our channel selection method yields less accuracy drop \mbox{(-0.15\%)} than~\cite{Tokovarov2020} \mbox{(-0.92\%)} using 14 \gls{eeg} channels. The authors did not further reduce the number of channels.}
Compared to~\cite{Dose2018AnBCIs,Wang2020_memea}, our method consistently achieves better accuracy at lower resource requirements. 
More specifically, a 14-channel configuration achieves 5.98\% better accuracy with 38.5$\times$ fewer parameters, 19.1$\times$ smaller feature map size, and 16.9$\times$ less computation than~\cite{Dose2018AnBCIs}.
\new{Compared to the method based on Granger causality proposed in~\cite{VARSEHI2021}, our solution is overall comparable, with slightly better accuracy when fewer channels are selected. %However, the mathematical operations (e.g., the calculation of the logarithm) required in~\cite{VARSEHI2021} are less hardware-friendly than our method.
With our method based on the spatial filters, we can decrease the number of \gls{eeg} channels down to 10, i.e., by a factor of 6.4$\times$, and still achieve comparable accuracy to the \gls{soa} with full-channel configuration, proving the effectiveness of our methods in extracting relevant features for the underlying task.} Compared to the baseline \edgeeegnet{} with 64 channels, using only 10 channels brings the advantage of 1.3$\times$ fewer parameters, 3.1$\times$ less memory requirement, and 1.4$\times$ less complexity in terms of \glspl{macc}, while retaining a similar classification accuracy of 82.51\%.
%82.51\% accuracy, which is only 0.28\% drop from the baseline with 64 \gls{eeg} channels. This configuration also yields 1.3$\times$ less parameters, 3.1$\times$ less memory requirement, and 1.4$\times$ less complexity in terms of \glspl{macc} compared to our full-channel model.

\new{We inspect the spatial filters of the trained models to gain insights into the brain activations. The average $\ell_2$-norm values of the spatial filters' weights are depicted in Fig.~\ref{fig:bci:heatmaps} and Fig.~\ref{fig:physio:heatmaps}. %Darker red colors represent higher values of the $\ell_2$-norm, meaning that the networks have learned stronger spatial weights for those brain regions. 
Each subject of the \bcicompivtwoa{} dataset presents a different distribution over the scalp. Overall, the higher values are near the electrodes C3, C4 over the sensorimotor cortex and from the Cz to POz over the temporal lobe, indicating that the networks have learned stronger spatial weights for these regions. The analysis on the \physionetmmmi{} dataset provides subject-independent, global information on the relevant brain regions. 
When considering the left- and right-hand tasks, the regions under the C3 and C4 have higher values. %They are responsible for hand movements~\cite{Pfurtscheller1999Event-relatedPrinciples}. 
The addition of the `rest' class introduces stronger learned weights around Iz over the occipital lobe, related to the visual cortex. The `feet' class induces activations around Cz over the sensorimotor cortex. These findings are confirmed in the literature~\cite{Pfurtscheller1999Event-relatedPrinciples,Zhao2019HandFootLip}. We additionally observe activations near the electrodes F7 and F8. This is likely due to the presentation of the visual cues. We discuss possible lines of investigation in Sec.~\ref{sec:discussion}.

%The accuracy values depicted in Fig. suggest that regions outside the sensorimotor cortex are also important for classifying the `rest' and `feet' classes, as the usage of distributed electrodes yields much better results than the headphone configuration, especially for the 3- and 4-class tasks. 
}

\subsection{Embedded Implementation}\label{sec:res:embedded}
\begin{table}[!b]
\setlength{\tabcolsep}{4.4pt}
\caption{\new{Networks deployment and measurement results.
%The average power consumption and the runtime per inference including the cluster startup time are measured. 
All Mr. Wolf cores run at 50\,MHz with 0.8\,V power supply as in~\cite{Schneider2020}. The \gls{soc} core of Vega runs at 50\,MHz and the cluster cores at 160\,MHz as in~\cite{Wang2021mrc_tbiocas}. The power supply is 0.65\,V.}} % OUT SHIFT not used
% (\gls{soc} + cluster domains)
\centering
\label{tab:netdeploy}
{
\begin{threeparttable}
\begin{tabular}{@{}lrrrrrr@{}}
\toprule
Dataset               & \multicolumn{3}{c}{BCI Comp.}  & \multicolumn{3}{c}{Physionet}  \\
\cmidrule(){1-1} \cmidrule(l){2-4} \cmidrule(l){5-7}
Num. classes          & 4            & 2         & 2         & 4               & 2              & 2 \\
Num. channels         & 22           & 22        & 6$^\dagger$         & 64              & 64             & 10$^\ddagger$ \\
(Q) Accuracy (\%)     & 75.63        & 86.52     & 97.76   & 65.31      & 82.61 & 82.51\\
Est. Memory (kB)           & 46.58        & 45.88     & 33.37    & 42.63  & 42.40           & 15.62 \\
%Est. \#\glspl{macc}        & 2209408      & 2209056         & 2208704   & 1824704   & 1505728         & 1505616      & 1505504        & 1090784 \\
Est. \#\glspl{macc} (M)       & 2.21      & 2.21   & 1.82   & 1.51         & 1.51        & 1.09 \\
Mr. Wolf \\
\hspace{1mm} Time/infer. (ms)   & {11.37}        & {11.30}     & {10.57}      & {6.21}            & {6.21}           & {5.53}    \\
\hspace{1mm} Avg. power (mW)       & {10.07}        & {10.24}     & {9.70}    & {9.92}            & {10.00}           & {9.06} \\
\hspace{1mm} Energy/infer. (\textmu J) & 114.5       & {115.7}    & {102.5}   & 61.60 & {62.10}         & {50.10} \\
\hspace{1mm} Throughput (MMACCs/s) & 194.3 & 195.5 & 172.6 & 242.5 & 242.4 & 197.2 \\
\hspace{1mm} En. Eff. (GMACCs/s/W) & 19.30 & 19.09 & 17.80 & 24.44 & 24.24 & 21.77 \\
% Throughput & \multirow{2}{*}{194.3} & \multirow{2}{*}{195.5} & \multirow{2}{*}{172.6} & \multirow{2}{*}{\textbf{242.5}} & \multirow{2}{*}{242.4} & \multirow{2}{*}{197.2} \\
% (MMACCs/s) & & & & & & \\
% %En. Eff. (GMACCs/s/W) & 22.85     & 22.85 & 22.45     & 21.18    & 28.73 & 28.97      & 28.24         & 27.05          \\
% En. Eff. & \multirow{2}{*}{19.30} & \multirow{2}{*}{19.09} & \multirow{2}{*}{17.80} & \multirow{2}{*}{\textbf{24.44}} & \multirow{2}{*}{24.24} & \multirow{2}{*}{21.77} \\
% (GMACCs/s/W) & & & & & & \\
Vega \\
\hspace{1mm} Time/infer. (ms)   & {5.10}        & {4.85}     & {4.73}      & {3.19}            & {3.13}           & \textbf{2.95}    \\
\hspace{1mm} Avg. power (mW)       & {11.87}        & {12.07}     & {11.52}    & {11.40}            & {11.28}           & {10.17} \\
\hspace{1mm} Energy/infer. (\textmu J) & 60.5       & {58.5}    & {54.5}   & 36.4 & {35.3}         & \textbf{30.0} \\
\hspace{1mm} Throughput (MMACCs/s) & 433.3 & 455.7 & 384.8 & 473.4 & \textbf{482.4} & 369.5 \\
\hspace{1mm} En. Eff. (GMACCs/s/W) & 36.50 & 37.75 & 33.40 & 41.52 & \textbf{42.77} & 36.33 \\
\bottomrule
\end{tabular}
\begin{tablenotes}\footnotesize
\item $^\dagger$ Subject 8. Selected channels: CP3, P1, POz, CP4, C6, FC4. % avg 25 runs
\item $^\ddagger$ Fold 1. Sel. channels: AF8, F8, T8, C3, Cz, C2, CP2, CP5, F6, T9. % check if it's from fold 1 and if averaged over 5 repetitions.
\end{tablenotes}
  \end{threeparttable}
  }
\end{table}

\new{We use the PyTorch-based Quantlab framework~\cite{quantlab2019} to quantize the models before the embedded deployment.
As described in Sec.~\ref{sec:embedded_impl}, we perform quantization-aware training with \gls{ste} and \gls{rpr} algorithms. For the \bcicompivtwoa{} dataset, $t_a$=450, $t_w$=550, and $t_{end}$=650 yield the best results.
We use the cross-entropy loss and the Adam optimizer with a fixed learning rate of 0.001 and eps=1e-7.
For the \physionetmmmi{} dataset, we have $t_a$=60, $t_w$=160, and $t_{end}$=260. For the \mbox{3-} and 4-class tasks, a fixed learning rate of 0.001 and eps=1e-9 yield the best results. Whereas for the 2-class task, the same learning rate scheduler as for the full-precision models is used with eps=1e-9.}
%
%\new{The training and the validation procedures are implemented in PyTorch based on the Quantlab framework~\cite{quantlab2019}.}

%We deploy the models with full-channel configuration for all three tasks and the ones with the lowest number of channels and the highest accuracy for the 2-class task to demonstrate the advantage of channel selection in reducing the energy consumption with experimental measurements.

%We measure the power consumption and the inference runtime of the deployed models using the Keysight N6705B power analyzer.

\begin{figure}
    \centering

  \resizebox{\columnwidth}{!}{%

% This file was created by tikzplotlib v0.9.9.
\begin{tikzpicture}

\begin{axis}[
name=myAxis,
width=\columnwidth, height=0.75\columnwidth,
legend cell align={center},
legend columns=3,
legend style={
  at={(0.5,0.99)},
  anchor=north,
  draw=none,
  font=\scriptsize
},
log basis x={10},
tick align=inside,
tick pos=left,
x grid style={black!20},
xlabel={Memory footprint [kB]},
xmajorgrids,
xmin=10, xmax=507847,
xmode=log,
%xtick style={color=black},
xlabel near ticks,
y grid style={black!20},
ylabel={Accuracy [\%]},
ymajorgrids,
ymin=62, ymax=90,
%ytick style={color=black},
ylabel near ticks,
clip=false
]

%networks = ['edgeEEGNet4(14)', 'edgeEEGNet3(11)', 'edgeEEGNet2(9)', 'edgeEEGNet3', 'edgeEEGNet2','edgeEEGNet','Q-EEGNet','EEGNet', 'S.ConvNet', 'FBCSP', 'MSFBCNN', 'EEG-TCNet', 'FB-3D-CNN']#, 'CNN++', 'TPCT']

% memfoot = [40.328, 37.678999999999995, 34.862, 46.231, 45.878, 46.584, 227.548, 910.192, 4241.2, 1244.0, 23720.0, 1601.08, 185895.2]
% accuracies = [75.84, 79.91, 86.21, 80.37, 86.32, 76.03, 70.9, 71.2, 74.31, 73.70, 75.80, 77.34, 86.96]#, 81.1, 88.87]
% maccs = [20.2, 19.5, 19, 22.1, 22.1, 22.1, 130,130, 630, 1040, 2020, 68, 623]#, 964, 17300] # in 0.1M
% maccs in M: edgeEEGNet 2.21; Q-EEGNet 13; S.ConvNet 63; FBCSP 104; MSFBCNN 202; EEG-TCNet 6.8; FB-3D-CNN 62.3

% for each 1M is 0.0025cm. therefore, to get the cm use the formula 0.1cm+X*0.0025cm with X the number of MACCs in M

\addplot[thick, draw=white!69.0196078431373!black, dashed, forget plot] coordinates {(64,62)(64,90)};
\node[] at (axis cs: 50,90.8) {\notsotiny 64\,kB};
\addplot[thick, draw=white!69.0196078431373!black, dashed, forget plot] coordinates {(512,62)(512,90)};
\node[] at (axis cs: 512,90.8) {\notsotiny 512\,kB};
\addplot[thick, draw=white!69.0196078431373!black, dash dot, forget plot] coordinates {(128,62)(128,90)};
\node[] at (axis cs: 150,90.8) {\notsotiny 128\,kB};
\addplot[thick, draw=white!69.0196078431373!black, dash dot, forget plot] coordinates {(1500,62)(1500,90)};
\node[] at (axis cs: 2000,90.8) {\notsotiny 1500\,kB};

\addlegendimage{color=colorgreen, mark=square*, only marks}
\addlegendentry{2-class}
\addlegendimage{color=colorblue, mark=square*, only marks}
\addlegendentry{3-class}
\addlegendimage{color=colorred, mark=square*, only marks}
\addlegendentry{4-class}

\fill[white!90!black] (4000,66.8) circle (0.1025cm) node[above=0.1025cm, yshift=-0.5ex, black] {\scriptsize 1\,M};
\fill[white!90!black] (10000,66.8) circle (0.225cm) node[above=0.225cm, yshift=-0.5ex, black] {\scriptsize 50\,M};
\fill[white!90!black] (32000,66.8) circle (0.35cm) node[above=0.35cm, yshift=-0.5ex, black] {\scriptsize 100\,M};
%\fill[white!90!black] (160000,66.8) circle (0.475cm) node[above, yshift=-0.5ex, black] {\scriptsize 150\,M};
\fill[white!90!black] (170000,66.8) circle (0.6cm) node[above=0.6cm, yshift=-0.5ex, black] {\scriptsize 200\,M};
% for each 1M is 0.0025cm. therefore, to get the cm use the formula 0.1cm+X*0.0025cm with X the number of MACCs in M

\fill[colorblue] (46.231,80.37) circle (0.105cm) node[above=0.105cm, yshift=-0.5ex, black] {\notsotiny \textbf{\edgeeegnet{}}};
\addplot[thick, gray, only marks, mark=*, mark size=0.6] coordinates {(46.231,80.37)(46.231,80.37)};
\fill[colorgreen] (45.878,86.32) circle (0.105cm) node[above=0.105cm, yshift=-0.5ex, black] {\notsotiny \textbf{\edgeeegnet{}}};
\addplot[thick, gray, only marks, mark=*, mark size=0.6] coordinates {(45.878,86.32)(45.878,86.32)};
\fill[colorred] (46.584,76.03) circle (0.105cm) node[above=0.105cm, yshift=-0.5ex, black] {\notsotiny \textbf{\edgeeegnet{}}};
\addplot[thick, gray, only marks, mark=*, mark size=0.6] coordinates {(46.584,76.03)(46.584,76.03)};

\fill[colorred] (68.15,70.9) circle (0.1325cm) node[below=0.1325cm, xshift=-1ex, yshift=0.5ex, black] {\notsotiny Q-\eegnet{}'20~\cite{Schneider2020}}; %227.548
\addplot[thick, gray, only marks, mark=*, mark size=0.6] coordinates {(68.15,70.9)(68.15,70.9)}; %227.548
\fill[colorred] (910.192,71.3) circle (0.1325cm) node[below=0.1325cm, xshift=2ex, yshift=0.5ex, black] {\notsotiny \eegnet{}'18~\cite{Lawhern2018EEGNet:Interfaces}};
\addplot[thick, gray, only marks, mark=*, mark size=0.6] coordinates {(910.192,71.3)(910.192,71.3)};
\fill[colorred] (4241.2,74.31) circle (0.2575cm) node[below=0.2575cm, xshift=0.5ex, yshift=0.5ex, black] {\notsotiny S. ConvNet'17~\cite{Schirrmeister2017DeepVisualization}};
\addplot[thick, gray, only marks, mark=*, mark size=0.6] coordinates {(4241.2,74.31)(4241.2,74.31)};
\fill[colorred] (23720,75.80) circle (0.605cm) node[above=0.605cm, xshift=1ex, yshift=-0.5ex, black] {\notsotiny MSFBCNN'19~\cite{wu2019_MSFBCNN}};
\addplot[thick, gray, only marks, mark=*, mark size=0.6] coordinates {(23720,75.80)(23720,75.80)};
\fill[colorred] (1601.08,77.35) circle (0.117cm) node[above=0.117cm, yshift=-0.5ex, black] {\notsotiny EEG-TCNet'20~\cite{ingolfsson2020eegtcnet}};
\addplot[thick, gray, only marks, mark=*, mark size=0.6] coordinates {(1601.08,77.35)(1601.08,77.35)};
\fill[colorred] (318,76.74) circle (0.1184cm) node[below=0.1184cm, yshift=0.5ex, black] {\notsotiny EEG-ITNet'22~\cite{Salami2022_eegitnet}};
\addplot[thick, gray, only marks, mark=*, mark size=0.6] coordinates {(318,76.74)(318,76.74)};
\fill[colorred] (58296,75.7) circle (0.3575cm) node[above=0.3575cm, yshift=-0.5ex, xshift=3.5ex, black] {\notsotiny MCNN'19~\cite{Amin2019_MCNN}};
\addplot[thick, gray, only marks, mark=*, mark size=0.6] coordinates {(58296,75.7)(58296,75.7)};

\fill[colorgreen] (185895.2,86.96) circle (0.25575cm) node[below=0.25575cm, xshift=-3ex, yshift=0.5ex, black] {\notsotiny FB3DCNN'21~\cite{bang2021fb3dcnn}};
\addplot[thick, gray, only marks, mark=*, mark size=0.6] coordinates {(185895.2,86.96)(185895.2,86.96)};

\fill[colorred] (74,73.72) rectangle (96,74.48) node[below, xshift=-0.5ex, yshift=-0.5ex, black] {\notsotiny MRC'21~\cite{Wang2021mrc_iscas}};
\addplot[thick, gray, only marks, mark=*, mark size=0.6] coordinates {(84,74.1)(84,74.1)};

    \end{axis}
    
  \def\myShift{-0.55cm}
  \draw [colorgreen, thick, ->] ([xshift=\myShift, yshift=1cm]myAxis.west) -- ([xshift=\myShift]myAxis.north west) node [midway, rotate=90, fill=white, yshift=2pt] {\footnotesize better} ;
  \def\myyShift{-0.57cm}
  \draw [colorgreen, thick, ->] ([yshift=\myyShift, xshift=-1.5cm]myAxis.south) -- ([yshift=\myyShift]myAxis.south west) node [midway, rotate=0, fill=white, xshift=2pt] {\footnotesize better} ;
  
  \draw [colorgreen, thick, <-] ([yshift=0.3cm, xshift=0.3cm]myAxis.south) -- ([yshift=0.3cm, xshift=-1.3cm]myAxis.south east) node [midway, rotate=0, fill=white, xshift=2pt] {\footnotesize better} ;

\end{tikzpicture}
}
    \caption{\new{Accuracy vs. memory footprint on the \bcicompivtwoa{} dataset. The number of \glspl{macc} is represented as the size of the circles, while the square indicates that it is variable. \xia{The memory footprint includes both parameters and features (see Table~\ref{tab:background}). The precision is assumed to be 32 bits if no quantization is proposed.} The dashed and dash-dotted lines mark the L1 and L2 memory of Mr. Wolf and Vega, respectively.}}% The smaller the circle size, the fewer the number of operations. The closer to the left top corner, the higher the accuracy and the lower the memory footprint.}
    \label{fig:accmemcomp}
\end{figure}
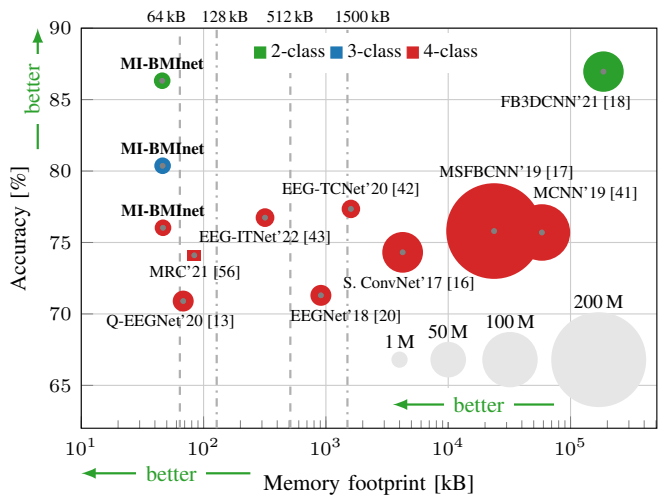

%
\new{The experimental results show that the quantized accuracy values are similar to the full-precision ones, with a maximum accuracy loss of 0.4\%. More precisely, the values for 2-, 3-, 4-class tasks are 86.52\%, 80.05\%, 75.63\% (\bcicompivtwoa{}), and 82.61\%, 75.12\%, 65.31\% (\physionetmmmi{}), respectively.}
%shows the comparison between the full-precision and the quantized models. In all cases, the accuracy loss is less than 1\%. 
The quantization of both weights and activations allows 4$\times$ reduction of the total memory footprint.
More specifically, the memory requirement is reduced from around 186\,kB and 171\,kB when using 32-bit representation down to roughly 47\,kB and 43\,kB with 8-bit quantization for the \bcicompivtwoa{} and \physionetmmmi{} datasets, respectively. 
Recall that the fast L1 memory of Mr. Wolf is only 64\,kB~\cite{pullini2019wolf}. 
\new{Hence, with a memory footprint of less than 50\,kB, we effectively eliminate the data transfer between the L1 and L2 memory during the computation of a single layer, reducing \lucar{double-buffering overheads} while maintaining similar classification accuracy.
%As illustrated in Fig.~\ref{fig:accmemcomp}, most related works do not fit in the fast memory of Mr. Wolf and Vega when considering both the total number of parameters and the maximum number of consecutive features and especially without quantization. %The only exceptions are EEGNet and \gls{mrc}; however, our model achieves a higher accuracy with fewer resources.
%Fig.~\ref{fig:accmemcomp} illustrates the accuracy versus the memory footprint (both the total number of parameters and the maximum number of consecutive features) and the computational requirements.
As illustrated in Fig.~\ref{fig:accmemcomp}, most related works do not fit in the fast memory of Mr. Wolf and Vega, especially without quantization.
The memory requirement of the quantized \edgeeegnet{} is up to four orders of magnitude lower, and it is within the \lucar{tightest} memory constraint (64\,kB).
%the memory availability of the L1 fast memory of the compute cluster of Mr. Wolf.
It also requires the least amount of \glspl{macc}.}
%and achieves similar accuracy as the related works.
%The \gls{simd} operations can also be used to speed up the execution by a factor of 4$\times$, while the 8-core compute cluster of Mr. Wolf is activated on demand to compute a classification inference.

\new{We measure the power consumption and the inference runtime of the deployed models using the Keysight N6705B power analyzer (61.44\,\textmu s sampling interval), including all operation domains of the selected microprocessors and the cluster startup time. We set the clock frequency of the cores to the same values as in~\cite{Schneider2020} and \cite{Wang2021mrc_tbiocas} for a fair comparison.
Table~\ref{tab:netdeploy} shows the results. 
The measured inference runtime on Mr. Wolf is around 11.4\,ms for the \bcicompivtwoa{} dataset in full-channel configuration and 6.2\,ms for the \physionetmmmi{} dataset. The average power consumption is roughly 10\,mW, yielding an energy consumption per inference of 115\,\textmu J and 62\,\textmu J, respectively. The execution time on Vega is more than twice faster and consumes down to 36\,\textmu J. Fig.~\ref{fig:measurements} depicts the comparison with the related works. Our implementation is orders of magnitude faster and less energy-hungry. Compared to the most energy-efficient model with the same \gls{mcu} configurations~\cite{Wang2021mrc_tbiocas}, our solution is 3.3$\times$ faster and more energy-efficient and achieves higher classification accuracy.
The channel reduction further reduces the runtime and the energy consumption.}
%The consumption is further reduced thanks to channel selection, as illustrated in Fig.~\ref{fig:measurements}.
%Considering the lowest number of selected channels, i.e., 4 channels for S5 of the \bcicompivtwoa{} dataset on 2-class task, the energy reduction during inference is \todo{XX\%} compared to the usage of all 22 channels.
\new{Considering the best performing subject (S8, 98.27\% accuracy) of the \bcicompivtwoa{} dataset on the 2-class task, the number of \gls{eeg} channels can be decreased to 6 while keeping similar classification accuracy (97.76\%). It yields an energy reduction of approximately 10\% compared to the usage of all 22 channels.
We can draw the same conclusion for the subject-independent \physionetmmmi{} dataset. 
We successfully decrease the number of channels by a factor of 6.4$\times$ (from 64 to 10) without significant accuracy loss for the 2-class task.
%For the 2-class task, the number of channels is reduced by a factor of 6.4$\times$, i.e., from 64 down to 10, without significant accuracy loss. 
It translates into a runtime speedup of 1.1$\times$ and an energy reduction of 15\%, i.e., down to 30\,\textmu J. Overall, the highest achieved throughput is 482.4\,M\glspl{macc}/s with an energy efficiency of 42.77\,G\glspl{macc}/s/W.}
%
%Compared to the related works, our proposed implementation is orders of magnitude faster and less energy-hungry, as shown in Fig.~\ref{fig:measurements}.

\input{02_figures/F_energyvsruntime}

\section{Discussion}\label{sec:discussion}

\new{
The proposed \gls{cnn} achieves similar accuracy as the most recent related works~\cite{ingolfsson2020eegtcnet,bang2021fb3dcnn,Salami2022_eegitnet,Tokovarov2020}, while being significantly more hardware-friendly.
%The authors in~\cite{ingolfsson2020eegtcnet} obtain a slightly higher classification accuracy. 
Note that our accuracy values are averaged over multiple runs to account for the variability caused by the non-deterministic behavior of the training algorithms for a better statistical significance. %Whereas, the accuracy reported in~\cite{ingolfsson2020eegtcnet} is the maximum value over multiple runs to obtain a better result. 
%Hence, our achieved accuracy (76.03\%) is fairly similar to the fixed models (77.35\%) proposed by~\cite{ingolfsson2020eegtcnet}.
The standard deviation among the training repetitions on the \bcicompivtwoa{} dataset is at least 1.3\%~\cite{Schneider2020}. Hence, our achieved accuracy is comparable to the related works.
The additional technique of subject-specific hyperparameters tunings proposed in~\cite{ingolfsson2020eegtcnet} has improved the accuracy to almost 84\%. \lucar{This further strengthens the evidence} that subject-specific network architectures (e.g., variable kernel sizes, different number of filters) lead to significant accuracy gains. This technique can be easily adapted to the proposed \gls{cnn}, and similar improvements can be expected. 

Notwithstanding, \edgeeegnet{} has roughly 10$\times$ less memory footprint and 3$\times$ fewer computations than~\cite{ingolfsson2020eegtcnet}, enabling its deployment on ultra-low-power \glspl{mcu} with real-time inference. The extraction of spatial information before the temporal features substantially reduces resource usage. It is a universal and effective technique that all related networks in Table~\ref{tab:background} can adopt to lower their resource demands, especially the recent ones with multiple parallel convolutional blocks~\cite{wu2019_MSFBCNN,Amin2019_MCNN,Salami2022_eegitnet}. Future directions can investigate subject-specific neural architecture search while being aware of the memory and computation burdens related to the order of the operations.
EEG-ITNet requires also relatively small resource usage thanks to the downsampling of the input \gls{eeg} signals~\cite{Salami2022_eegitnet}. It is also an effective way to reduce the model size as demonstrated in~\cite{Wang2020_memea}, \xia{and it can be considered as a preprocessing stage in addition to our methods.}

The proposed channel reduction algorithm further reduces the resource requirements while retaining similar accuracy. It additionally lowers the power consumption of the data acquisition stage.
\lucar{Considering the same setup as in~\cite{Kartsch2019BioWolf:Connectivity_short}, i.e., 0.965\,mW/channel, 1.26\,mW for the digital section excluding the processing, and a 65\,mAh battery, the reduction from 64 to 10 channels yields an increased battery life from 3.8 to 22 hours with the negligible consumption of 0.03\,mW using Vega or 0.05\,mW using Mr. Wolf, assuming a continuous classification every second.}
%The extremely low energy per inference of the proposed solutions yields a negligible consumption of 0.01\,mW using Vega or 0.017\,mW using Mr. Wolf with a continuous classification of 3\,s non-overlapping window, without degrading the battery life and maintaining it at 22 hours. 
%This negligible consumption does not degrade the battery life maintaining it at 22\,h.} %Unlike the related works~\cite{belwafi2018_wolacsp,Wang2020_memea}, the energy per inference of our solution does not degrade the battery life even with a continuous operation.

The inspection of the spatial filters gives insights into the relevant brain regions for \gls{mi} tasks.
Especially interesting are the activations around F7 and F8 electrodes. They cover the inferior frontal gyrus near the Sylvian fissure between the frontal and temporal lobe, where the insular cortex is located~\cite{HomanCerebralLocationEEG1020sys,TOWLE19931SylvianFissure}. This area is considered relevant in attentional elaborations and working memory processing related to external stimuli~\cite{Tops2011InferiorFrontalGyrus,Corbetta2008Environment}. 
Being part of the dorsal frontoparietal network, it has a top-down control function over motor and sensory areas with the basic cognitive selection of sensory information and response~\cite{Corbetta2002stimulusAttention}. Moreover, a study suggests that it plays a role in motor planning and imagery~\cite{PTAK2017DorsalFrontoparietalNetworkMotor}. A similar line of research on the insular cortex, saccade system, and supplementary motor areas further confirms the involvement of this area in motor control, attentional fixation, and responses to switching stimuli~\cite{Anderson1994InsularCortexMotorControl,nachev2008supplimentarymotorareas}. 
%These pieces of evidence lead to the conclusion that the activation of this area is likely due to the presentation of the cue and the experimental protocol of the \physionetmmmi{} dataset. However, 
Another study suggests that the activities in this area are due to electromyographic and electrooculographic artifacts~\cite{Dose2018AnBCIs}. Depending on the acquisition setup, this is also a valid hypothesis since the electrodes F7 and F8 are prone to collect muscular and ocular artifacts~\cite{Sazgar2019eegartifacts}.
%Similar heatmaps activations are also reported by Tokovarov et al.~\cite{Tokovarov2020} for 2-class task for the same dataset.
% cit paper F7 inferior frontal gyrus, 
%brodman area 6 which contains premotor cortex cit nature paper
% cit paper There is considerable variation in the distance
% between the fMRI-defined primary motor and language cortex
% and the most contiguous electrode.
% pfurtscheller
% primary motor cortex, premotor cortex, and supplementary motor cortex.
%
An acquisition procedure that eliminates the dependency on the visual cue is helpful in future research to investigate the fundamental nature of this activation. A possible solution is to design an experimental paradigm combined with electromyograms~\cite{xu2014ankle-foot}.

Another future research direction is the detection of movement intention. For a real-world \gls{iom} scenario, no visual cue is presented to the subject marking the \gls{mi}'s start. Hence, algorithms that can autonomously detect the onset of the \gls{mi} intention are necessary for an online \gls{bmi} that is asynchronously self-paced. %Finally, a full system that is asynchronously self-paced and capable of detecting and classifying different \gls{mi} tasks has to be designed for the future \gls{iom}.

}

\section{Conclusion}\label{sec:conclusion}

This paper proposes an energy-efficient embedded solution for \gls{mi}-\glspl{bmi}. We design a tiny \gls{cnn} that achieves similar \gls{soa} accuracy but is orders of magnitude less resource-demanding.
%and can operate with a variable number of \gls{eeg} channels for application-specific selection. 
We further propose an automatic channel reduction method based on spatial filters and extract the most relevant \gls{eeg} channels to effectively reduce the memory requirements, the computational complexity, and the power consumption. 
%
%The optimal numbers of channels for subject-specific models on BCI Competition IV-2a dataset are 9, 11, and 14 on average across the subjects for \mbox{2-}, \mbox{3-}, and \mbox{4-class} tasks, respectively. Whereas for the subject-independent models on Physionet \gls{mmmi} dataset the number of channels can be reduced down to 10, 20, and 18, respectively, as a trade-off between accuracy and energy-efficiency.
%
\new{We finally deploy the proposed models on ultra-low-power \glspl{mcu} and experimentally measure the runtime and the energy consumption. The final solution consumes down to 30\,\textmu J and takes only 2.95\,ms per inference with an accuracy of 82.51\% on the 2-class \gls{mi} task using only 10 instead of 64 \gls{eeg} channels, yielding an operation of up to 22 hours. 
%We successfully satisfy the three-way trade-off among accuracy, resource cost, and power usage and set the new \gls{soa} for embedded \gls{mi}-\glspl{bmi}.
%
By combining a model design that is aware of the resource usage, the channel reduction and the quantization techniques, and an embedded implementation that optimally exploits the underlying hardware architecture, we satisfy the three-way trade-off among accuracy performance, resource usage, and power consumption, setting the new SoA for next-generation wearable BMIs with smart edge computing.
}

%In this work, we presented the full workflow from the hardware/software codesign of a new \gls{cnn} that is resource-friendly yet accurate, to the further optimizations using channel selection, and finally an energy-efficient deployment on a ultra-low power \gls{mcu} for the future generation of smart \gls{mi}-\gls{bmi} wearable device.
%
% We propose an embedded model based on EEGNet for low-power \gls{mi}-\glspl{bci}. The proposed model achieves 2.05\%, 5.25\%, and 6.49\% higher accuracy than the \gls{soa} CNN on \mbox{2-,} \mbox{3-,} and 4-class \gls{mi} classification, while requiring two orders of magnitude less memory for storing model parameters and 4.6$\times$ less memory for feature maps during inference execution. 
% % 
% We reduce the input feature map by downsampling in the temporal and spatial domain as well as narrowing down the time window and relax the memory requirements by 7.6$\times$ at 0.31\% accuracy loss, and by 15$\times$ at 2.51\% loss. 
% % 
% We demonstrate the performance of the proposed models on two commercial \glspl{mcu}. In particular, the implemented models execute in around 44\,ms consuming 18.1\,mJ per inference on an ARM Cortex-M7 and in 101\,ms using 4.28\,mJ on an ARM Cortex-M4F processor, making them suitable for a battery-operated real-time wearable system to continuously perform online MI classification. 
%

% \section*{Acknowledgment}
% This project was supported by the Swiss Data Science Center PhD Fellowship under grant ID P18-04.

\bibliographystyle{IEEEtran}
\bibliography{ref_michael,bib,IEEEtranBSTCTL}{}
% check the following stackexchange for shortening bib to use et al.
% https://tex.stackexchange.com/questions/16506/bibliography-author-initial-spacing
% i.e., change the setting in IEEEtranBSTCTL.bib!
% the same for use url

\begin{IEEEbiography}[{\includegraphics[width=1in,height=1.25in,clip,keepaspectratio]{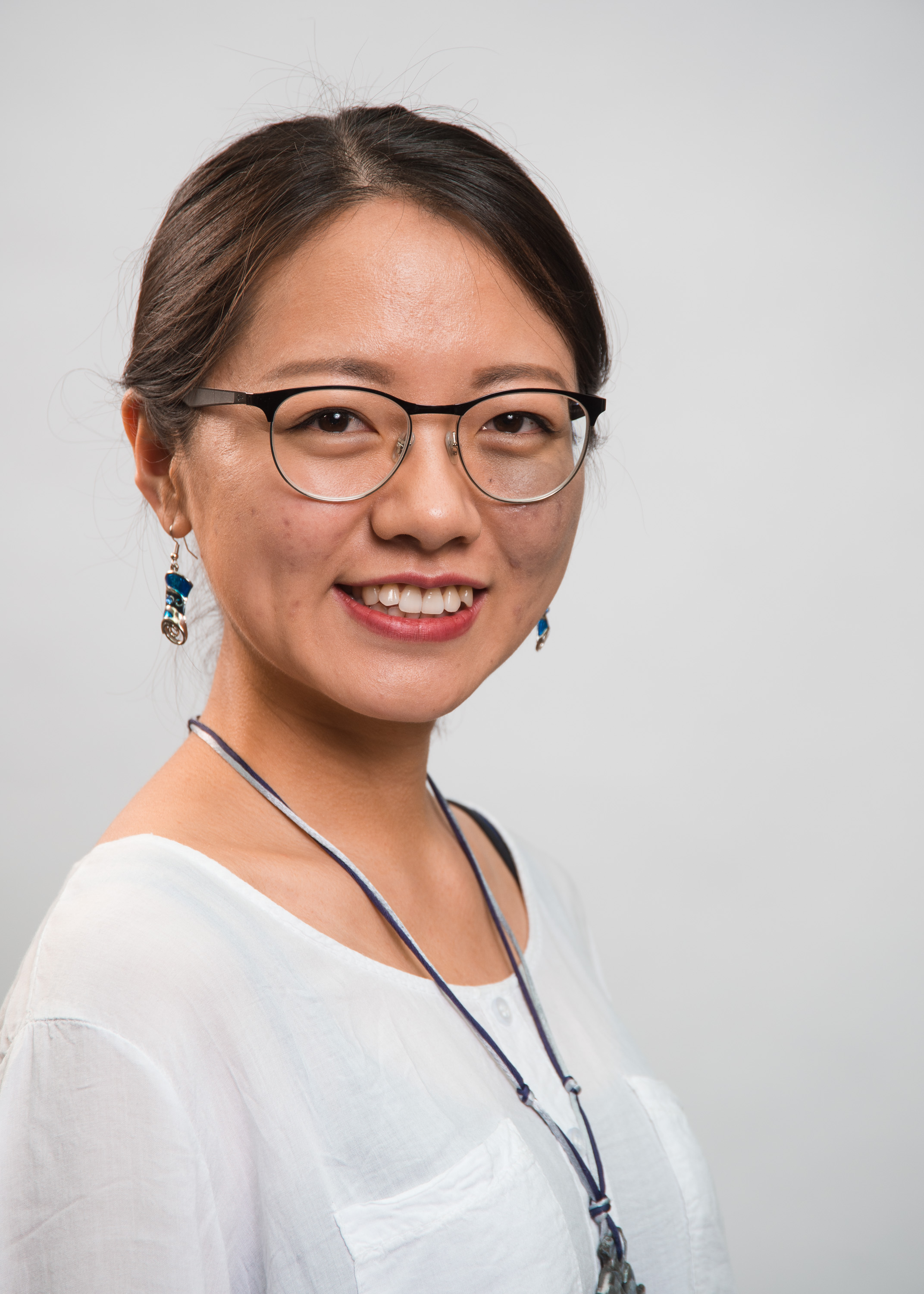}}]{Xiaying Wang}
received her B.Sc. and M.Sc. degrees in biomedical engineering from Politecnico di Milano, Italy, and ETH Zürich, Switzerland, in 2016 and 2018, respectively. She is currently pursuing a Ph.D. degree at the Integrated Systems Laboratory at ETH Zürich. Her research interests include biosignal processing, brain--machine interface, low power embedded systems, energy-efficient smart sensors, and machine learning on microcontrollers. She received the excellent paper award at the IEEE Healthcom conference in 2018 and won the Ph.D. Fellowship funded by the Swiss Data Science Center in 2019. 
\end{IEEEbiography}

\begin{IEEEbiography}[{\includegraphics[width=1in,height=1.25in,clip,keepaspectratio]{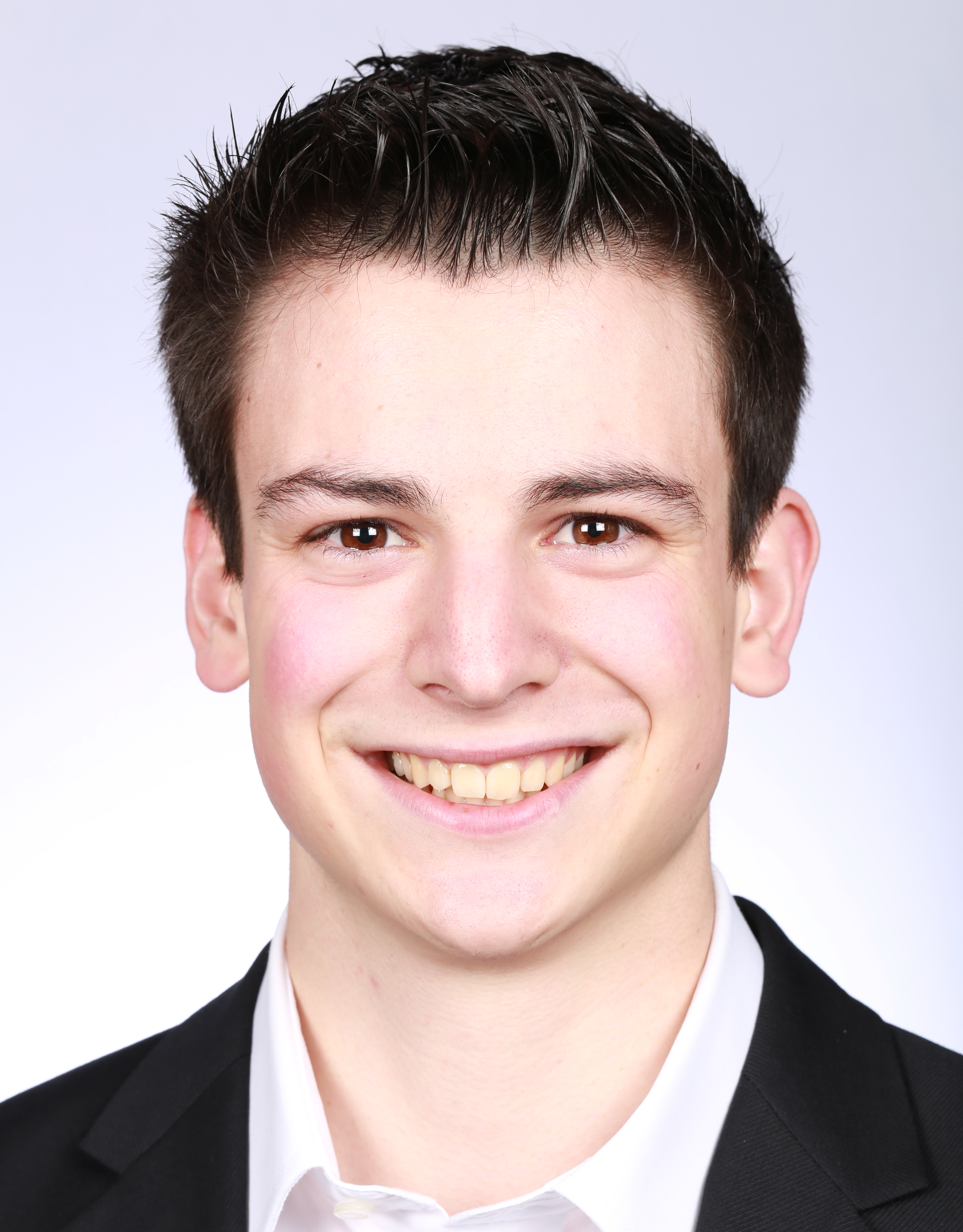}}]{Michael Hersche}
received his M.Sc. degree from the Swiss Federal Institute of Technology Zurich (ETHZ), Switzerland, where he is currently pursuing a Ph.D. degree. Since 2019, he has been a Research Assistant with ETHZ in the group of Prof. L. Benini at the Integrated Systems Laboratory. His research interests include digital signal processing, artificial intelligence, and communications with focus on hyperdimensional computing. He received the 2020 IBM Ph.D. Fellowship Award.
\end{IEEEbiography}

\begin{IEEEbiography}[{\includegraphics[width=1in,height=1.25in,clip,keepaspectratio]{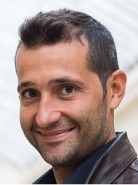}}]{Michele Magno}
received his master’s and Ph.D. degrees in electronic engineering from the University of Bologna, Bologna, Italy, in 2004 and 2010, respectively. He is currently a Senior Researcher and lecturer with ETH Zürich, Switzerland. He has authored more than 150 papers in international journals and conferences, a few of them awarded as best papers. His current research interests include wireless sensor networks, wearable devices, energy harvesting, low power management techniques, and extension of the lifetime of batteries-operating devices. 
\end{IEEEbiography}

\begin{IEEEbiography}[{\includegraphics[width=1in,height=1.25in,clip,keepaspectratio]{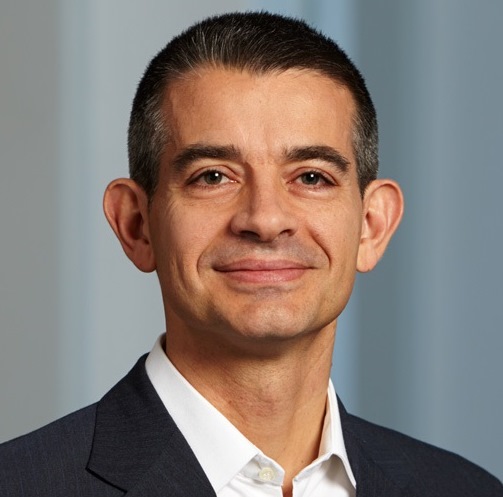}}]{Luca Benini}
is the Chair of Digital Circuits and Systems at ETH Zürich and a Full Professor at the University of Bologna. He has served as Chief Architect for the Platform2012 in STMicroelectronics, Grenoble. Dr. Benini’s research interests are in energy-efficient systems and multi-core SoC design. He is also active in the area of energy-efficient smart sensors and sensor networks. He has published more than 1’000 papers in peer-reviewed international journals and conferences, four books, and several book chapters. He is a Fellow of the ACM and the IEEE and a member of the Academia Europaea.
\end{IEEEbiography}

\clearpage

\section*{Appendix}

\noindent
\begin{minipage}{\textwidth}
  \begin{minipage}[b]{0.49\textwidth}
    \centering
    \fontsize{5}{7}\selectfont
    \includegraphics[width=0.9\columnwidth]{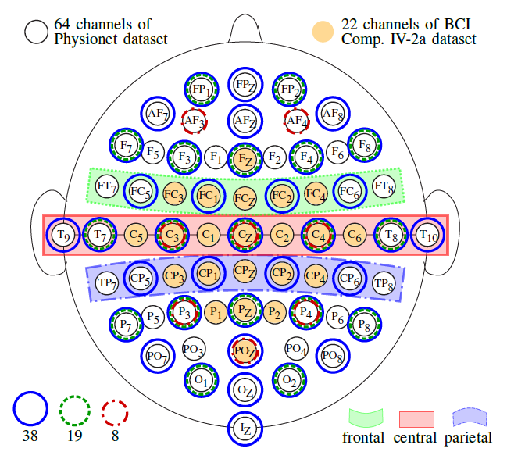}
    \captionof{figure}{Electrode configurations for manual channel selections.}
    \label{fig:channels}
  \end{minipage}
  \hfill
  \begin{minipage}[b]{0.49\textwidth}
    \centering
    \captionof{table}{EEG electrodes configurations using headset-based channel selection over sensorimotor and neighboring areas. The regions correspond to the brain areas with color-shaded background illustrated in Fig.~\ref{fig:channels}, as well as the electrodes, while $N_{ch}$ is the number of selected channels.}
    \label{tab:headset_channels}
    
% \usepackage{booktabs}
% \usepackage{multirow}
%\newcolumntype{C}[1]{@{}>{\columncolor{white}[0pt]\centering\arraybackslash}p{#1}@{}}

% \begin{table}[!t]
% 	\centering
% 	\caption{EEG electrodes configurations using headset-based channel selection over sensorimotor and neighboring areas. The regions correspond to the brain areas with color-shaded background illustrated in Fig.~\ref{fig:channels}, as well as the electrodes, while $N_{ch}$ is the number of selected channels.}
% 	\label{tab:headset_channels}
\footnotesize
\begin{tabular}[b]{lcr}
\toprule
Region                         & $N_{ch}$                                     & Electrodes                                                                                                                                                                              \\ \midrule
                              & 2                                            & C3, C4                                                                                                                                                                                  \\
                              & \cellcolor[HTML]{EFEFEF}3                    & \cellcolor[HTML]{EFEFEF}C3, CZ, C4                                                                                                                                                      \\
                              & 5                                            & C5, C3, CZ, C4, C6                                                                                                                                                                      \\
                              & \cellcolor[HTML]{EFEFEF}7                    & \cellcolor[HTML]{EFEFEF}C5, C3, C1, CZ, C2, C4, C6                                                                                                                                      \\
                              & 9                                            & T7,C5, C3, C1, CZ, C2, C4, C6,T8                                                                                                                                                        \\
\multirow{-6}{*}{Central (C)}  & \cellcolor[HTML]{EFEFEF}11                   & \cellcolor[HTML]{EFEFEF}T9,T7,C5,C3,C1,CZ,C2,C4,C6,T8,T10                                                                                                                               \\ \midrule
                              & 4                                            & C3, C4, FC3, FC4                                                                                                                                                                        \\
                              & \cellcolor[HTML]{EFEFEF}6                    & \cellcolor[HTML]{EFEFEF}C3, CZ, C4, FC3, FCZ, FC4                                                                                                                                       \\
                              & 10                                           & C5, C3, CZ, C4, C6, FC5,FC3, FCZ, FC4,FC6                                                                                                                                               \\
                              & \cellcolor[HTML]{EFEFEF}                     & \cellcolor[HTML]{EFEFEF}                                                                                                                                                                \\
                              & \multirow{-2}{*}{\cellcolor[HTML]{EFEFEF}14} & \multirow{-2}{*}{\cellcolor[HTML]{EFEFEF}\begin{tabular}[c]{@{}r@{}}C5, C3, C1, CZ, C2, C4, C6, \\ FC5, FC3, FC1, FCZ, FC2, FC4, FC6\end{tabular}}                                      \\
                              &                                              & T7,C5, C3, C1, CZ, C2, C4, C6,T8,                                                                                                                                                       \\
                              & \multirow{-2}{*}{18}                         & \multicolumn{1}{l}{FT7, FC5, FC3, FC1, FCZ, FC2, FC4, FC6, FT8}                                                                                                                         \\
                              & \cellcolor[HTML]{EFEFEF}                     & \cellcolor[HTML]{EFEFEF}                                                                                                                                                                \\
\multirow{-9}{*}{C+Frontal}  & \multirow{-2}{*}{\cellcolor[HTML]{EFEFEF}20} & \multirow{-2}{*}{\cellcolor[HTML]{EFEFEF}\begin{tabular}[c]{@{}r@{}}T9,T7,C5,C3,C1,CZ,C2,C4,C6,T8,T10, \\ FT7, FC5, FC3, FC1, FCZ, FC2, FC4, FC6, FT8\end{tabular}}                     \\ \midrule
                              & 4                                            & C3, C4, CP3, CP4                                                                                                                                                                        \\
                              & \cellcolor[HTML]{EFEFEF}6                    & \cellcolor[HTML]{EFEFEF}C3, CZ, C4, CP3, CPZ, CP4                                                                                                                                       \\
                              & 10                                           & C5, C3, CZ, C4, C6, CP5, CP3, CPZ, CP4, CP6                                                                                                                                             \\
                              & \cellcolor[HTML]{EFEFEF}                     & \cellcolor[HTML]{EFEFEF}                                                                                                                                                                \\
                              & \multirow{-2}{*}{\cellcolor[HTML]{EFEFEF}14} & \multirow{-2}{*}{\cellcolor[HTML]{EFEFEF}\begin{tabular}[c]{@{}r@{}}C5, C3, C1, CZ, C2, C4, C6, \\ CP5, CP3, CP1, CPZ, CP2, CP4, CP6\end{tabular}}                                      \\
                              &                                              &                                                                                                                                                                                         \\
                              & \multirow{-2}{*}{18}                         & \multirow{-2}{*}{\begin{tabular}[c]{@{}r@{}}T7,C5, C3, C1, CZ, C2, C4, C6,T8, TP7, \\ TP7, CP5, CP3, CP1, CPZ, CP2, CP4, CP6, TP8\end{tabular}}                                         \\
                              & \cellcolor[HTML]{EFEFEF}                     & \multicolumn{1}{l}{\cellcolor[HTML]{EFEFEF}}                                                                                                                                            \\
\multirow{-9}{*}{C+Parietal} & \multirow{-2}{*}{\cellcolor[HTML]{EFEFEF}20} & \multicolumn{1}{l}{\multirow{-2}{*}{\cellcolor[HTML]{EFEFEF}\begin{tabular}[c]{@{}r@{}}T9,T7,C5,C3,C1,CZ,C2,C4,C6,T8,T10,\\  TP7, CP5, CP3, CP1, CPZ, CP2, CP4, CP6, TP8\end{tabular}}} \\ \bottomrule
\end{tabular}
  \end{minipage}
  \vfill
  \vspace{1cm}
  \begin{minipage}{\textwidth}
    \centering
    \captionof{table}{\xia{Classification accuracy (\%)\,/\,kappa score of \edgeeegnet{} with all channels and with the minimum number of selected channels (in brackets) when allowing a maximum accuracy degradation of 1\% on the IV-2a dataset.}}
    \label{tab:bciRes_1perc}
    % \begin{table*}[!b]
\setlength{\tabcolsep}{2.6pt}
%  \rotatebox{90}{%
%   \begin{minipage}{0.9\textheight}
%\captionsetup{width=3in}
% \caption{\new{Classification accuracy (\%)\,/\,kappa score of \edgeeegnet{} with all channels and with the minimum number of selected channels (in brackets) when allowing a maximum accuracy degradation of 1\% on the IV-2a dataset.}}
% \centering
% \label{tab:bciRes_1perc}
%\footnotesize
\begin{tabular}{@{}lrrrrrr@{}}
\toprule
        & \multicolumn{2}{c}{2-class}                                                                  & \multicolumn{2}{c}{3-class}                  & \multicolumn{2}{c}{4-class}\\
\cmidrule(lr){2-3} \cmidrule(lr){4-5} \cmidrule(lr){6-7}
S. & all ch. & sel. ch. & all ch. & sel. ch. & all ch. & sel. ch. \\
\midrule
1       & 84.03\,/\,0.68      & 86.98\,/\,0.74\,(4)        & 89.43\,/\,0.84      & 91.03\,/\,0.87\,(6)       & 83.10\,/\,0.78      & 83.89\,/\,0.79\,(14) \\
2       & 71.15\,/\,0.42      & 72.03\,/\,0.44\,(16)       & 65.74\,/\,0.49      & 68.76\,/\,0.53\,(16)       & 59.27\,/\,0.46      & 59.52\,/\,0.46\,(16) \\
3       & 94.95\,/\,0.90      & 94.10\,/\,0.88\,(4)       & {91.67\,/\,0.88}      & 91.65\,/\,0.88\,(16)       & {90.64\,/\,0.88}      & 90.97\,/\,0.88\,(18) \\
4       & 74.38\,/\,0.49      & 76.07\,/\,0.52\,(4)        & 75.66\,/\,0.64      & 76.00\,/\,0.64\,(14)       & 69.77\,/\,0.60      & 70.63\,/\,0.61\,(18) \\
5       & 92.00\,/\,0.84      & 93.84\,/\,0.88\,(4)        & 79.17\,/\,0.70      & 79.54\,/\,0.69\,(10)       & 71.83\,/\,0.62      & 72.64\,/\,0.64\,(10) \\
6       & 79.48\,/\,0.59      & {81.11}\,/\,0.62\,(20)     & 63.48\,/\,0.45      & 64.69\,/\,0.47\,(16)      & 58.10\,/\,0.44      & 59.42\,/\,0.46\,(4) \\
7       & 90.51\,/\,0.81      & 90.63\,/\,0.81\,(14)     & 88.49\,/\,0.83      & 88.57\,/\,0.83\,(16)     & 84.71\,/\,0.80      & 84.90\,/\,0.80\,(16) \\
8       & {98.06\,/\,0.96}      & 97.76\,/\,0.96\,(4)        & 88.18\,/\,0.82      & 88.26\,/\,0.82\,(10)      & 84.55\,/\,0.80 & 85.93\,/\,0.81\,(16) \\
9       & 92.31\,/\,0.85      & 92.62\,/\,0.85\,(10)       & 81.53\,/\,0.72      & 81.33\,/\,0.72\,(19)       & 82.33\,/\,0.76 & 81.67\,/\,0.76\,(16) \\
\cmidrule(lr){2-3} \cmidrule(lr){4-5} \cmidrule(lr){6-7}
Avg.    & 86.32\,/\,0.73      & 87.24\,/\,0.75\,(8.9)   & 80.37\,/\,0.71      & 81.09\,/\,0.72\,(13.7)    & 76.03\,/\,0.68      & 76.62\,/\,0.69\,(14.2) \\
Std.   & 8.96\,/\,0.18        & 8.40\,/\,0.17\,(6)     & 9.78\,/\,0.15        & 9.23\,/\,0.14\,(3.9)      & 11.11\,/\,0.15       & 10.95\,/\,0.15\,(4.3) \\ 
\bottomrule
\end{tabular}
% \end{minipage}}
% \end{table*}
  \end{minipage}
\end{minipage}

% \begin{figure}[!t]
%     \centering
%     \fontsize{5}{7}\selectfont
%     %\includegraphics[width=\linewidth, trim={0cm 2.4cm 0cm 6cm}, clip=true]{02_figures/physionet_bcicomp_headch.pdf} %ch_head_physio.pdf}
%     \includesvg[width=0.9\columnwidth]{02_figures/physionet_bcicomp_headch.svg}
%     \caption{Electrode configurations for manual channel selections. %The circled ones are to compare with~\cite{Wang2020_memea}, while the ones on color-shaded background are for analyses over sensorimotor and neighboring regions.
%     }
%     \label{fig:channels}
% %    \vspace{-0.4cm}
% \end{figure}

% \input{02_figures/T_headset_channels}

%\input{02_figures/T_iv2aResCS_1perc}

%\input{02_figures/F_heatmaps_bci}

% \bibliographystyle{IEEEtran}
% \bibliography{ref_michael,bib,IEEEtranBSTCTL}{}

\end{document}